\documentclass[a4paper,11pt]{article}

\usepackage{amsmath}
\usepackage{bm}
\usepackage{graphicx}
\usepackage{comment}

\begin{document}

%
%
%
%
%
%

\title{Quantum-statistical equation-of-state models: high-pressure Hugoniot shock
adiabats}

\author{J.C. Pain\footnote{Commissariat \`a l'\'{E}nergie Atomique, CEA/DIF, Bo\^ite Postale 12, 91680 Bruy\`{e}res-le-Ch\^{a}tel Cedex, France}}

\maketitle


\begin{abstract}
We present a detailed comparison of two self-consistent equation-of-state models
which differ from their electronic contribution: the atom in a
spherical cell and the atom in a jellium of charges. It is shown that both
models are well suited for the calculation of Hugoniot shock adiabats in the
high pressure range (1 Mbar-10 Gbar), and that the atom-in-a-jellium model
provides a better treatment of pressure ionization. Comparisons with
experimental data are also presented. Shell effects on shock adiabats are
reviewed in the light of these models. They lead to
additional features not only in the variations of pressure versus density, but
also in the variations of shock velocity versus particle velocity. Moreover,
such effects
are found to be responsible for enhancement of the electronic specific
heat.
\end{abstract}


\section{Introduction}

The studies concerning matter under extreme conditions have broad applications
to material science, inertial confinement fusion and astrophysics. About 98 \%
of the Universe is made of hot
and ionized matter. The center of stars and white dwarfs have pressures
exceeding thousands of Mbar and temperatures of tens of millions of K. In
neutron stars pressures of millions of Mbar are believed to exist. 
 
Nowadays, the shock-wave technique in laboratory
laser experiments \cite{rem} enables one to reach maximum pressures that are as
high as
hundreds of
megabars and even more for many materials \cite{nel0}. Such pressures range to
three orders
of magnitude higher than the pressure in the Earth center and close to that in
the center of the Sun.

Therefore, the need for suitable equation of state (EOS) of high-energy-density
matter becomes crucial. The thermodynamics and the hydrodynamics of these
systems can not be predicted without a knowledge of the EOS which describes how
a material reacts to pressure. For instance, the theory of stellar evolution is
affected by uncertainties in EOS (the internal structure and cooling time of the
white dwarfs depend on the
details of their EOS). More precisely, in astrophysical objects such as low-mass
stars, brown dwarfs and giant planets, hydrogen and helium highly dominate, and
their EOS, including non-ideality effects, can be accurately evaluated using
``chemical'' free-energy models (see for instance \cite{win,jur1,jur2}).
However, the EOS of carbon is also required for modeling inner envelopes of
carbon-rich white dwarfs or outer accreted envelopes of neutron stars
\cite{cha0}. For general astrophysical applications of various types of stars,
the EOS of oxygen is also required. In the center of the Sun, where matter
density is about 157 g/cm$^3$ and temperature 1.3 keV, although helium and
hydrogen are the most abundant elements, the small quantity of iron has a strong
impact on the calculation of the radiative opacity. Therefore, it is important
to be able to determine the partial density of iron, which requires knowledge of
its EOS.

In the
present work, we consider strongly coupled (nonideal) plasmas, characterized by
a high density and/or a low temperature. In such plasmas, ions
are strongly correlated, electrons are partially degenerate, the De Broglie
wavelength of the electron becomes of the same order of the interparticle
distance, and the coupling parameter 

\begin{equation}\label{defgam}
\Gamma=\Big(\frac{4\pi}{3}\Big)^{1/3}\frac{Z^{*2}n_i^{1/3}}{k_BT},
\end{equation}

ratio of Coulomb potential energy and thermal energy, is much larger than
unity. The matter density $n_i$ represents the number of particles per volume
unit, $Z^*$ the average ionization of the plasma, $k_B$ the Boltzmann constant
and $T$ the temperature (throughout this paper, physical quantities are
expressed in atomic units (a.u.)). 
 
When a material is subjected to a strong shock wave, it becomes compressed,
heated and ionized.
As the strength of the shock is varied for a fixed initial state, the
pressure-density final states of the material behind the shock belongs to a
curve named shock adiabat or Hugoniot curve. The
Hugoniot curve depends on the EOS of the matter, which, in principle, can be
determined from theory. In practice, the barriers to \textit{ab initio}
calculations are formidable owing to the computational difficulty of solving the
many-body problem. Consequently, it has proven necessary to introduce
simplifying approximations into the governing equations. The price of
simplification, however, is typically a loss of generality and the resulting
theory is adequate only over a limited region of the phase diagram. The
determination of the average electronic charge density relies usually on the
Density Functional Theory. A well-known example is the Thomas-Fermi (TF) model
of dense matter \cite{fey}, which contains certain essential features in order
to characterize the material properties over a limited range of conditions. At
intermediate shock pressures, when the material becomes partially ionized, the
EOS depends on the precise quantum-mechanical state of the matter, \textit{i.e.}
on the electronic shell structure \cite{cha0,des}. Therefore, there is a great
interest in the
physics of bound levels in high-energy-density plasmas  \cite{ebe} and quantum
self-consistent-field (QSCF) models are replacing the TF approach. Such
statistical models are well-suited for elements with a few
occupied orbitals, \textit{i.e.} with a ``reasonable'' number of electrons.
Therefore, the
theory presented in this paper is not applicable to hydrogen and helium, since
their atomic number is too low. 
A lot of
theoretical and computational efforts have been made in the United States and in
Russia in the last twenty years concerning quantum effects on Hugoniot shock
adiabats. It is important to review that huge amount of work, in the light of
the most recent developments in the modeling of atomic structure. Indeed, QSCF
 calculations strongly depend on the modeling of the ions
and on the treatment of free electrons. We present a QSCF-based EOS model, which
will be named ESODE (Equation of State with 
Orbital Description of Electrons) in the following. The ionic contribution to
the EOS is described
by a perfect ideal gas, and the cold curve (T=0 K isotherm) has been obtained in
most of the cases from Augmented Plane Wave (APW) calculations \cite{louc}.
Exchange-correlation effects at finite temperature are taken into account. ESODE
comes in two main versions for the thermal electronic contribution to the EOS:
Average atom in a Spherical Cell (ASC) or Average atom in a Jellium of Charges
(AJC). In both cases, bound electrons are treated quantum-mechanically. In the
ASC model, all electrons are confined within a Wigner-Seitz sphere, and the
continuity theorem of the wavefunctions at the boundary can be satisfied in
different ways. In the AJC model, bound-electron wavefunctions can extend
outside the sphere, where the plasma is represented by a uniform electronic
density (jellium) neutralized by a continuous background of positive charges,
representing ions. An interesting feature of AJC model is the possibility to use
a quantum-mechanical description of free states (instead of a semi-classical
one) together with a careful search for shape resonances. This seems to be
necessary in order to ensure thermodynamic consistency, and a suitable
description of polarization and density effects. The QSCF models (ASC and AJC)
are described in section \ref{sec2}. Shock adiabats calculated from those models
are presented, analyzed and compared to traditional TF model and to published
experimental data in section \ref{sec3}. An interpretation of shell effect in
terms of specific heat is discussed in
section \ref{sec3} as well. The dependence of shock velocity on particle
velocity
is analyzed in section \ref{sec4}.


\section{\label{sec2}The models}

\begin{figure}
\begin{center}
\includegraphics[width=8.6cm, trim=0 0 0 -60]{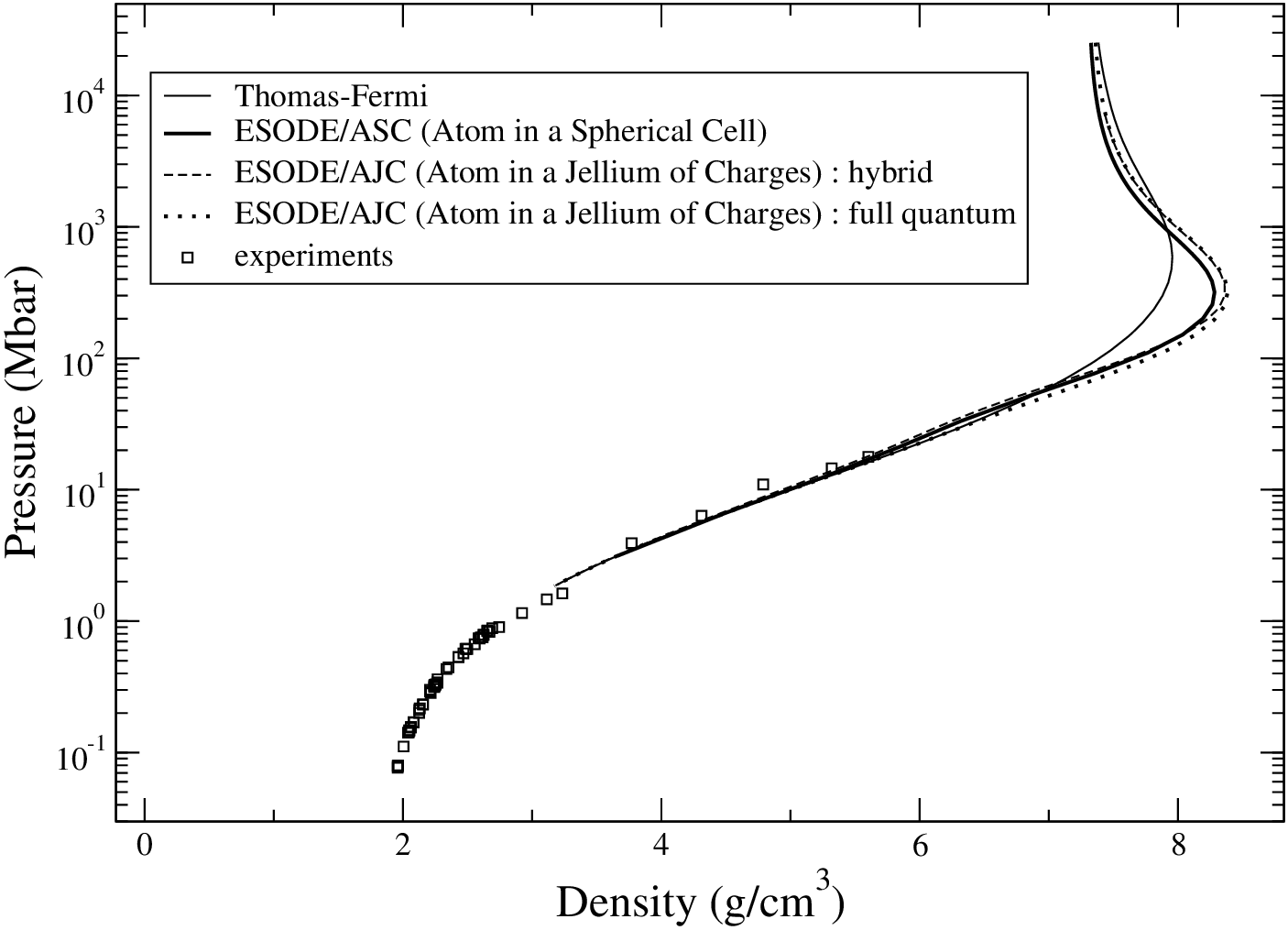}
\end{center}
\caption{Hugoniot curves for Be. $\rho_0=$1.83 g/cm$^3$. Experimental values are
given by Refs. \cite{rus, wal, isb, mar, rag, nel}.}
\label{fig1}
\end{figure}


\subsection{\label{subsec21}First quantum statistical model: Atom in a Spherical
Cell (ASC)}

The task is to evaluate the contribution to the EOS coming from the excitations 
of the electrons due to temperature and compression. Atoms in a plasma can be
idealized by an average atom confined in a Wigner-Seitz (WS) sphere, which
radius $r_{ws}$ is related to matter density. Inside the sphere, the electron
density has the following form \cite{roz2}:

\begin{eqnarray}\label{rel1}
n(r)&=&\sum_bf_l(\epsilon_b,\mu)\Big|\psi_b(\vec{r})\Big|^2\nonumber\\
& &+\frac{\sqrt{2}(k_BT)
^{3/2}}{\pi^2}[J_{1/2}\Big(-\bar{V}(r),\bar{\mu}-\bar{V}(
r),\chi\Big)\nonumber \\
& &+\chi J_{3/2}\Big(-\bar{V}(r),\bar{\mu}-\bar{V}(r),\chi\Big)],
\end{eqnarray}

where 

\begin{equation}
f_l(x,y)=\frac{2(2l+1)}{1+\exp[(x-y)/k_BT]}
\end{equation}

is the usual Fermi-Dirac population and 

\begin{equation}
\tilde{J}_{n/2}(a,x,\sigma)=\int_a^{\infty}\frac{y^{n/2}(1+\sigma
y/2)^{1/2}}{1+e^{y-x}}dy
\end{equation}

is the modified Fermi function of order $n/2$. The first term in (\ref{rel1})
corresponds to the contribution of bound electrons to the charge
density, while the second term is the free-electron contribution, written in its
semi-classical TF form. The energy $\epsilon_b$ and wavefunction $\psi_b$ of a
bound orbital are calculated in the Pauli approximation
\cite{pau}, in which only first-order relativistic corrections to the
Schr\"odinger equation 

\begin{equation}\label{sch1}
-\frac{1}{2}\Delta \psi_b+[V(r)+V_{mv}(r)+V_D(r)]\psi_b=\epsilon_b\psi_b
\end{equation}

are retained. We introduce the notations
$\chi=k_BT/E_0$, $E_0$ being the rest mass energy of the electron and
$\bar{V}(r)=V(r)/(k_BT)$, where $V(r)$ is the self-consistent potential:

\begin{equation}\label{scf2}
V(r)=-\frac{Z}{r}+\int_0^{r_{ws}}\frac{n(r')}{|\vec{r}-\vec{r'}|}d^3r'+V_{xc}(r),
\end{equation}

$V_{xc}$ being the exchange-correlation contribution, evaluated in the local
density approximation \cite{iye}. V$_{mv}$ is the mass-velocity correction 

\begin{equation}
V_{mv}(r)=-\frac{[\epsilon_b-V(r)]^2}{2E_0}
\end{equation}

and V$_D$ the Darwin correction

\begin{equation}
V_D(r)=-\frac{1}{2}\frac{1}{2E_0+\epsilon_b-V(r)}\frac{dV}{dr}\Big(\frac{d}{dr}-\frac{1}{r}\Big).
\end{equation}

Omission of the spin-orbit term is a consequence of
the spherical symmetry. Last, the chemical potential $\mu$ is obtained from the
neutrality of the ion sphere:

\begin{equation}\label{scf3}
\int_0^{r_{ws}}n(r)4\pi r^2dr=Z,
\end{equation}

and $\bar{\mu}=\mu/(k_BT)$. Eqs. (\ref{rel1}), (\ref{sch1}), (\ref{scf2}) and 
(\ref{scf3}) must be solved self-consistently provided that, at each step of the
iterative
process towards convergence, bound orbitals are obtained from the Schr\"odinger
equation. The electronic pressure $P_e$ \cite{pai,pai2} consists of three
contributions, $P_e=P_{b}+P_{f}+P_{xc}$, where the bound-electron pressure
$P_{b}$ is evaluated using the stress-tensor formula

\begin{eqnarray}\label{presb}
P_{b}&=&\sum_{b}\frac{f(\epsilon_b,\mu)}{8\pi
r_{ws}^2(1+\frac{\epsilon_b}{2E_0})}\Big[\Big(\frac{dy_b}{dr}\Big|_{r_{ws}}\Big
)^2\nonumber\\
& &+\Big(2\epsilon_b(1+\frac{\epsilon_b}{2E_0})-\frac{1+l+l^2}{r_{ws}^2}\Big)y_
b^2(r_{ws})\Big],
\end{eqnarray}

$y_b$ representing the radial part of the wavefunction $\psi_b$ multiplied by
$r$. The free-electron pressure $P_{f}$ reads

\begin{eqnarray}\label{presf}
P_{f}&=&\frac{2\sqrt{2}}{3\pi^2}(k_BT)^{5/2}[\tilde{J}_{3/2}(-\bar{V}(r_{ws}
),\bar{\mu}-\bar{V}(r_{ws}),\chi)\nonumber\\
& &+\frac{\chi}{2}\tilde{J}_{5/2}(-\bar{V}(r_{ws}),\bar{\mu}-\bar{V}(r_{ws}),
\chi)]
\end{eqnarray}

\begin{figure}
\begin{center}
\includegraphics[width=8.6cm,trim=0 0 0 -60]{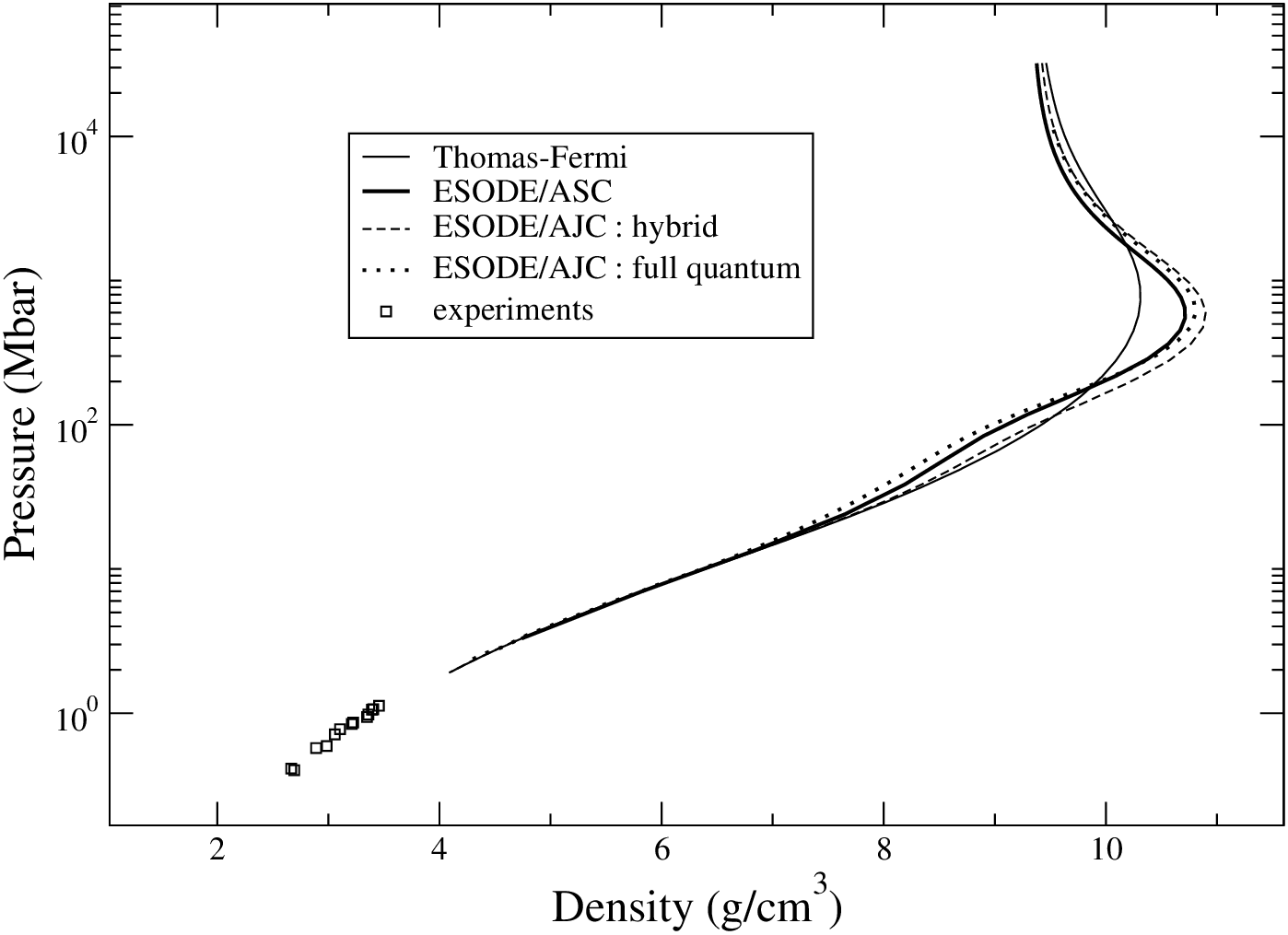}
\end{center}
\caption{Hugoniot curves for B. $\rho_0=$2.34 g/cm$^3$. Experimental values are
given by Ref. \cite{rus, mar}.}
\label{fig2}
\end{figure}

and $P_{xc}$ is the exchange-correlation pressure evaluated in the local density
approximation \cite{iye}. The choice of the boundary conditions plays a major
role in the expression of pressure. We chose a decreasing-exponential boundary
condition, since it allows the matching of wavefunctions
with Bessel functions outside the WS sphere \cite{pai}, where the potential is
zero. Moreover, such a condition is consistent with the fact that the
bound-electron density vanishes at infinity. The internal energy in the ASC
model is

\begin{figure}
\begin{center}
\includegraphics[width=8.6cm,trim=0 0 0 -60]{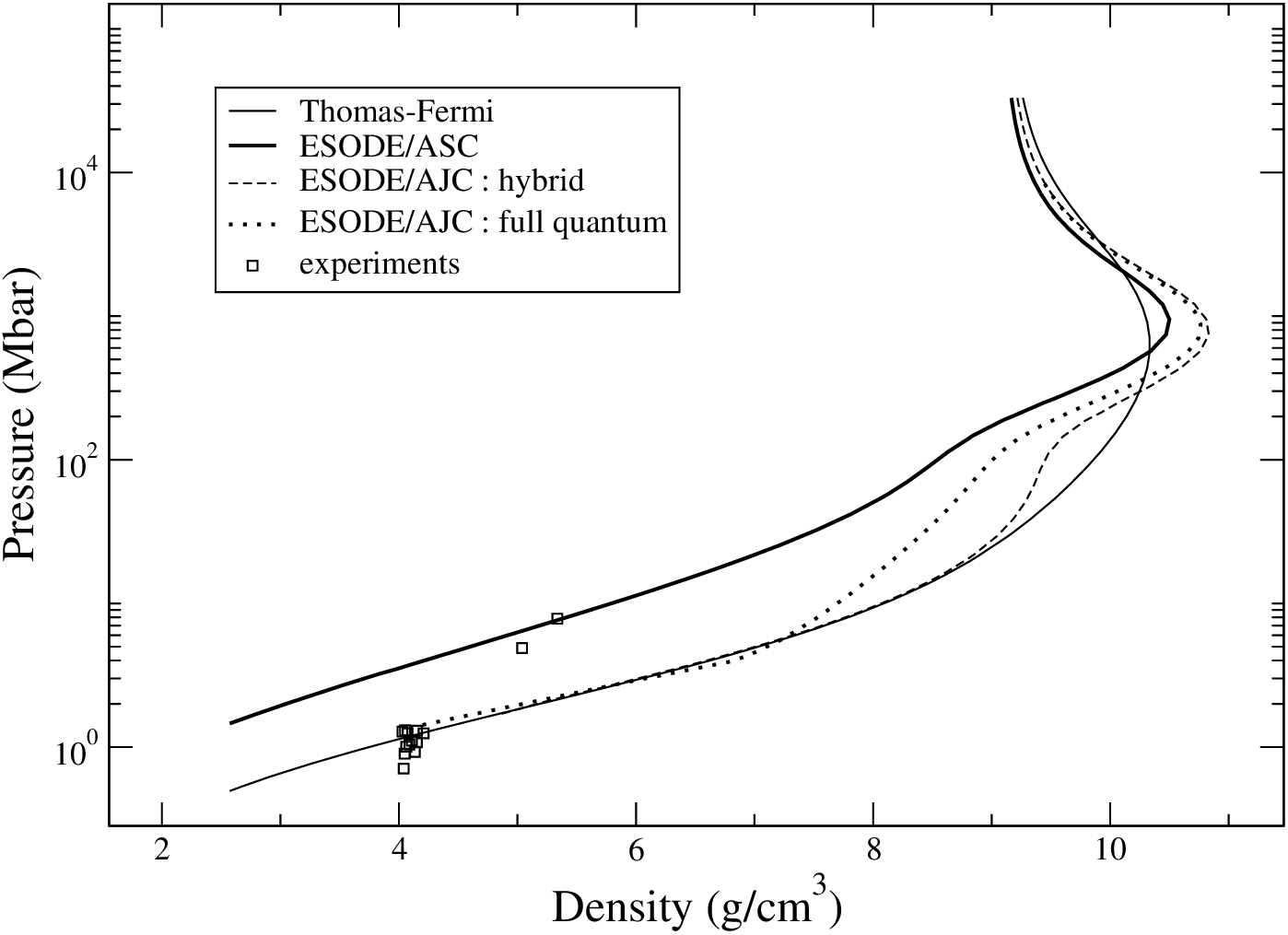}
\end{center}
\caption{Hugoniot curves for C (graphite). $\rho_0=$2.28 g/cm$^3$. Experimental
values are given by Ref. \cite{rus, mar, gus, nel2}.}
\label{fig3}
\end{figure}

\begin{eqnarray}\label{int}
E_e&=&\sum_jn_j\epsilon_j-\frac{1}{2}\int_0^{r_{ws}}n(r)\int_0^{r_{ws}}
\frac{n(r')}{|\vec{r}-\vec{r'}|}d^3rd^3r'\nonumber\\
& &+E_{xc}-\int_0^{r_{ws}} n(r)V_{xc}(n(r))d^3r,
\end{eqnarray}

where $E_{xc}$ is the exchange-correlation internal energy and $n_j$ the
population of state $j$ (either bound or free). The first term in (\ref{int})
can be expressed by

\begin{equation}
\sum_jn_j\epsilon_k=\sum_bf_l(\epsilon_b,\mu)\epsilon_b+E_{f,k}+E_{f,p}.
\end{equation}

$E_{f,p}$ is the potential energy

\begin{eqnarray}\label{relp}
E_{f,p}&=&\frac{\sqrt{2}(k_BT)^{3/2}}{\pi^2}\int_0^{r_{ws}}[\tilde{J}_{1/2}(
-\bar{V}(r),\bar{\mu}-\bar{V}(r),\chi)\nonumber\\
& &+\chi\tilde{J}_{3/2}(-\bar{V}(r),\bar{\mu}-\bar{V}(r),\chi)]
\bar{V}(r)d^3r
\end{eqnarray}

and $E_{f,k}$ the kinetic energy

\begin{eqnarray}\label{rel2}
E_{f,k}&=&\frac{\sqrt{2}(k_BT)^{5/2}}{\pi^2}\int_0^{r_{ws}}
[\tilde{J}_{3/2}(-\bar{V}(r),\bar{\mu}-\bar{V}(r),\chi)\nonumber\\
& &+\chi\tilde{J}_{5/2}(-\bar{V}(r),\bar{\mu}-\bar{V}(r),\chi)]d^3r.
\end{eqnarray}

\begin{figure}
\begin{center}
\includegraphics[width=8.6cm,trim=0 0 0 -60]{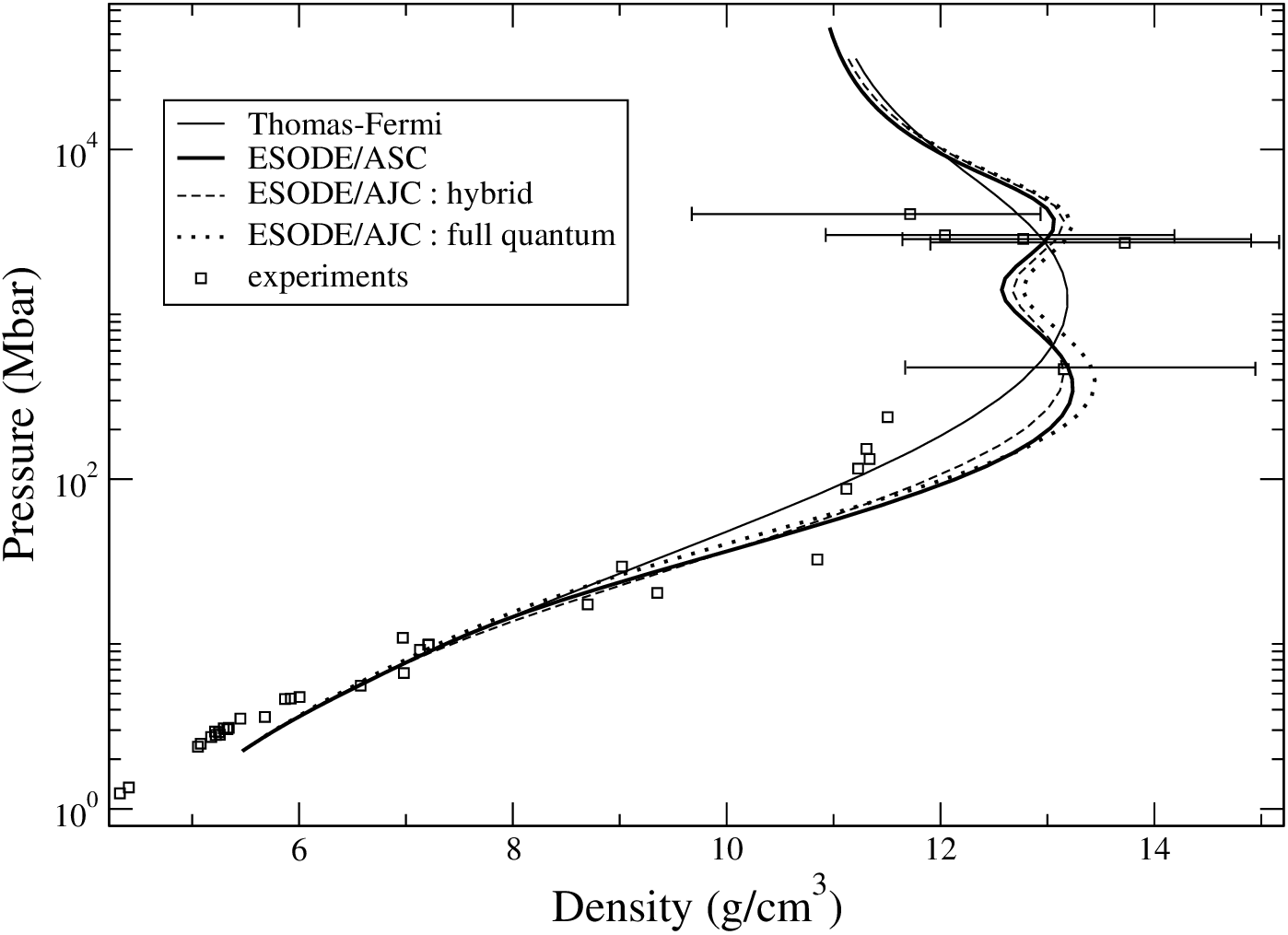}
\end{center}
\caption{Hugoniot curves for Al. $\rho_0=$2.70 g/cm$^3$. Experimental values are
given by Ref. \cite{rus, alt, vol, vla, rag2, sim1, avr, pod, tru2, knu}.}
\label{fig4}
\end{figure}

In the present work, only non-relativistic calculations are carried out, which
correspond to $E_0\rightarrow\infty$ in Eqs. (\ref{rel1}), (\ref{presb}),
(\ref{presf}), (\ref{relp}) and (\ref{rel2}), in order to make comparisons with
the model described below (in
section \ref{subsec22}), which is non-relativistic. Formulas (\ref{rel1}) to
(\ref{rel2}) enable one to include relativistic effects without solving Dirac
equation, in which the wavefunction consists of two components, making the
choice of boundary conditions even more awkward. 


\subsection{\label{subsec22}Second quantum statistical model: Atom in a Jellium
of Charges (AJC)}

In order to go beyond the TF approximate treatment of continuum electron charge
density, it is necessary to use a full quantum-mechanical description of the
continuum states and to consider that both bound and free orbitals can extend
outside the WS sphere. This leads us to define the environment outside and far
from the central average ion. One way to address this is the jellium model (or
electron gas model), \textit{i.e.} a uniform electron density $-\bar{n}_+$
neutralized by a positive background $\bar{n}_+$ simulating the ionic charges.
The model relies on a method proposed by J. Friedel \cite{fri1,fri2} in order to
treat the electronic structure of an impurity, represented by a spherical
potential of finite range, in an electron gas. It has been further developed by
L. Dagens \cite{dag1,dag2} and F. Perrot \cite{per1,per2,per3}. In this
framework, the electron density can be written:

\begin{equation}
n(r)=-\bar{n}_++\Delta n_a+\Delta n_{ab}+\Delta n_{abc}+...,
\end{equation}

\begin{figure}
\begin{center}
\includegraphics[width=8.6cm,trim=0 0 0 -60]{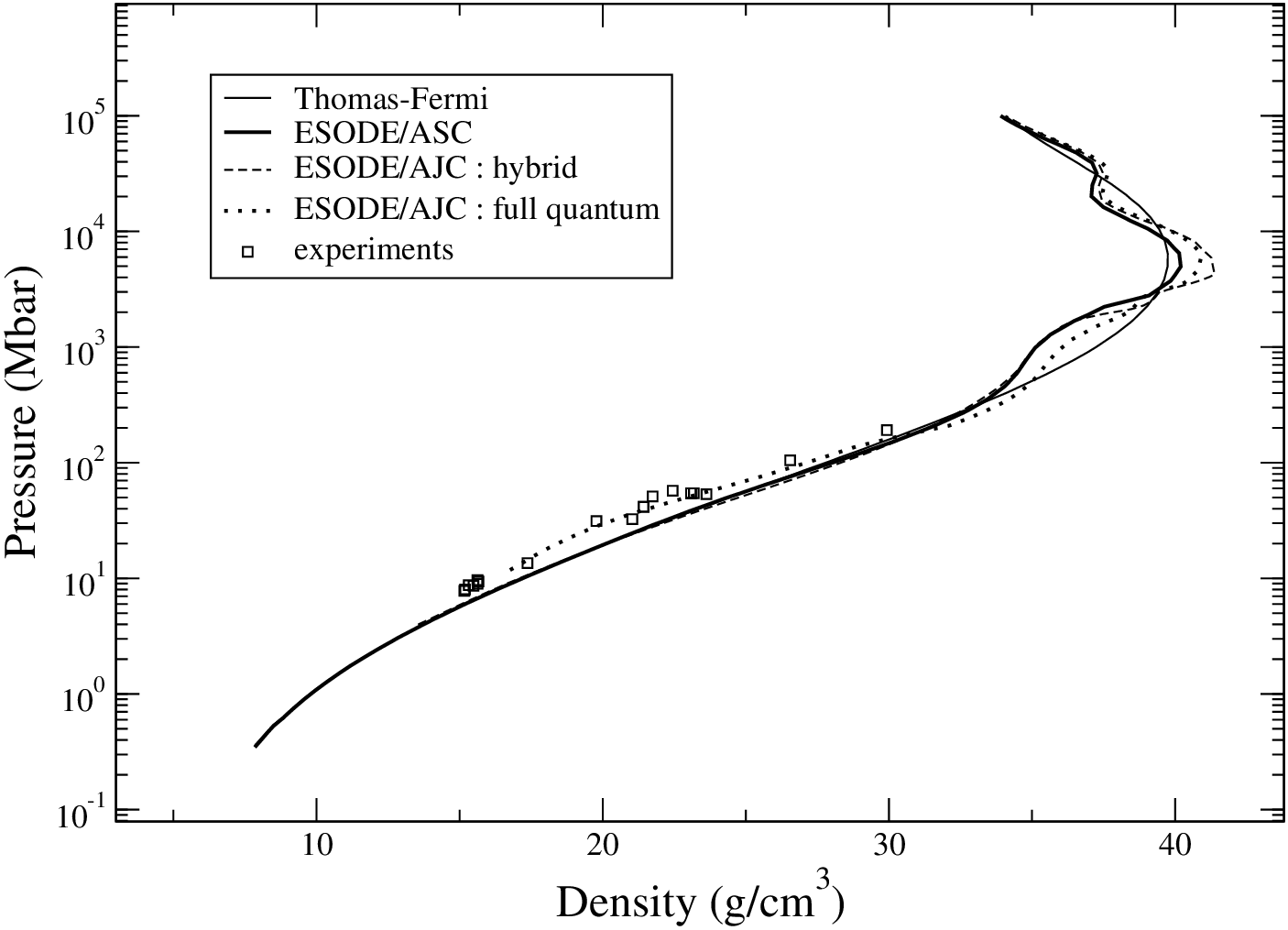}
\end{center}
\caption{Hugoniot curves for Fe. $\rho_0=$7.85 g/cm$^3$. Experimental values are
given by Ref. \cite{rus, alt, rag2, alt2, kru, alt3, tru3, alt4, tru4, tru5,
tru6,
alt5}.}
\label{fig5}
\end{figure}

where $-\bar{n}_+$ is the density of the electron gas (jellium), $\Delta n_a$
the charge density displaced by the immersion of an atom ``a'' in the jellium,
$\Delta n_{ab}$ the difference between the charge density of the $a-b$ molecule
and the charge densities of free atoms $a$ and $b$, \textit{etc}. The AJC
approach retains only the first two terms. A nucleus of charge $Z$ is then
introduced in a cavity in the positive background. The radius of the cavity
represents the average atomic radius in the plasma. Therefore, the problem is
reduced to the response of the electrons to the immersion of a positive charge
$Z$. Ionic response is roughly simulated by the formation of a cavity. These
considerations lead to the following form of the electron density:

\begin{equation}
n(r)=\sum_bf_l(\epsilon_b,\mu)\Big|\psi_b(\vec{r})\Big|^2\nonumber\\+\sum_l\int_0^{\infty}f
_l(\epsilon,\mu)\Big|\psi_{\epsilon,l}(\vec{r})\Big|^2d\epsilon,
\end{equation}

where the eigenstates $(\psi_b,\epsilon_b)$ and the scattering states
$(\psi_{\epsilon,l},\epsilon)$ are obtained solving Schr\"odinger equation with
the potential

\begin{equation}
V(r)=-\frac{Z}{r}\int_0^{R_{\infty}}\frac{n(r')}{|\vec{r}-\vec{r'}|}d^3r'+V_{xc}
(r)-V_{xc}(-\bar{n}_+),
\end{equation}

where $R_{\infty}>>r_a$, $r_a$ being the radius of the cavity. As in the ASC
model, $V_{xc}$ is calculated from \cite{iye}. $V(r)$ is determined, as in the
ASC model, in a self-consistent way. The ionic density is modeled by

\begin{equation}
n_+(r)=\left\{ \begin{array}{ll}
                   0      & \mbox{for $r<r_a$} \\
                   \bar{n}_+ & \mbox{for $r>r_a$.}
		   \end{array}\right.
\end{equation}

\begin{figure}
\begin{center}
\includegraphics[width=8.6cm,trim=0 0 0 -60]{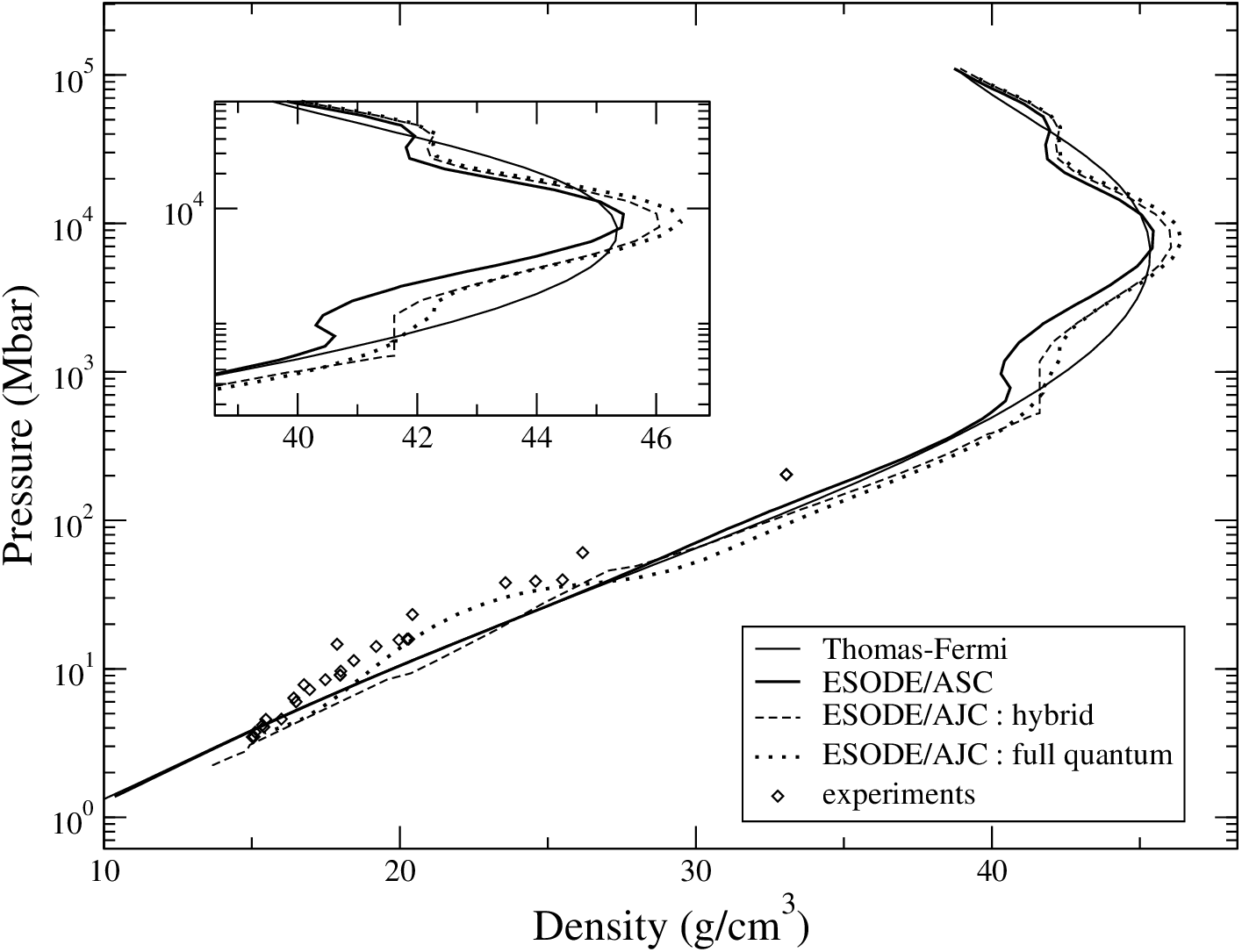}
\end{center}
\caption{Hugoniot curves for Cu. $\rho_0=$8.84 g/cm$^3$. Experimental values are
given by Ref. \cite{rus, isb, alt, rag2, alt2, tru6, alt5b, kor, tru7, tru8,
alt6,
glu, mit, tru9}.}
\label{fig6}
\end{figure}

The expression of internal energy is \cite{per3}

\begin{equation}
E[n,n_+]=Z ^*[e_k+e_{xc}](-\bar{n}_+)+\Delta E[n,n_+]-\Lambda_TP_1v_a,
\end{equation}

with $v_a=4\pi r_a^3/3$, $\Lambda_T=(\partial \ln Z^*/\partial \ln T)|_{v_a,T}$
and

\begin{equation}
P_1=\frac{\bar{n}_+}{v_a}\int_{r_a}^{R_{\infty}}4\pi r^2V(r)dr. 
\end{equation}

$e_k$ and $e_{xc}$ are the kinetic and exchange-correlation energies per free
electron and $\Delta E[n,n_+]$ is the energy change resulting from the immersion
of an ion in the jellium:

\begin{eqnarray}
\Delta
E[n,n_+]&=&\int_0^{R_{\infty}}(e_k[n(r)]-e_k[-\bar{n}_+])d^3r\nonumber\\
& &-\int_0^{R_{\infty}
}\frac{Z}{r}(n(r)+n_+(r))d^3r\nonumber\\
&
&+\frac{1}{2}\int_0^{R_{\infty}}\int_0^{R_{\infty}}\frac{n(r)+n_+(r)}{|\vec{r}-
\vec{r'}|}(n(r')+n_+(r'))d^3rd^3r'\nonumber\\
& &+\int_0^{R_{\infty}}(e_{xc}(n(r))-e_{xc}(-\bar{n}_+))d^3r.
\end{eqnarray}

The pressure is obtained by

\begin{equation}\label{pv}
P_e=[P_k+P_{xc}](-\bar{n}_+)+\bar{n}_+U(r_a)+(1-\Lambda_{v_a,T})P_1,
\end{equation}

where $\Lambda_{v_a}=(\partial \ln Z^*/\partial \ln {v})|_{v_a,T}$ and

\begin{eqnarray}
U(r)&=&-\frac{Z}{r}+\frac{1}{r}\int_0^r4\pi
r'^2(n(r')+n_+(r'))dr'\nonumber\\
& &+\int_r^{R_{\infty}}4\pi r'(n(r')+n_+(r'))dr'.
\end{eqnarray}

The term $P_k$ corresponds to the pressure of a free-electron gas. The
expression of pressure in (\ref{pv}) is rigorously the derivative of energy
versus volume. This is the main difference with INFERNO model \cite{lib,roz1}.
The major difficulty of these models is that the average ionization is not well
defined when the outer electrons are more or less delocalized. The only way to
make the formalism variational is to specify the ionization in the AJC model. In
other words, the question is how to define the residual electron density
$(-\bar{n}_+)$ far away from the point where the positive charge is introduced
into the jellium. A convenient choice is $Z^*(v)=Z^*_{TF}(v)$ \cite{per3},
$Z^*_{TF}$ being the TF ionization. In such a way derivatives of ionization with
respect to volume and temperature can be obtained analytically, using the
numerical fit proposed by R.M. More \cite{mor2,sal}.

ASC and AJC models are very different concerning the modeling of the
environment of the atom, isolated and confined in the ASC model, and immersed in
an infinite effective medium in the AJC model. 

\begin{figure}
\begin{center}
\includegraphics[width=8.6cm,trim=0 0 0 -60]{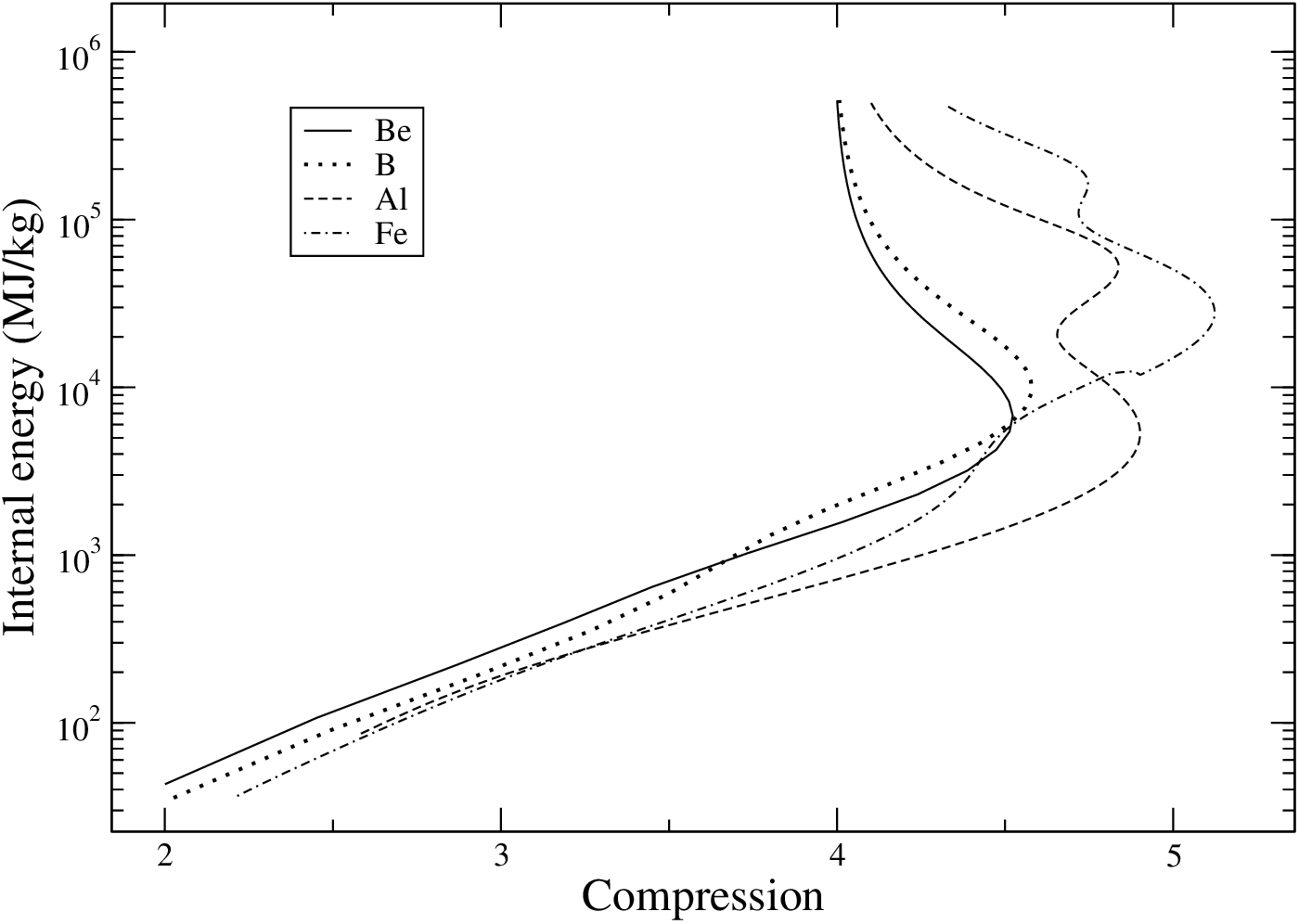}
\end{center}
\caption{Internal energy versus compression along the Hugoniot curve for Be, B,
Al and Fe in the ASC model.}
\label{fig7}
\end{figure}


\subsection{\label{subsec25}Nuclear contribution and cold curve}

The adiabatic approximation is used to separate the thermodynamic functions into
electron and nuclear (ionic) components. The total pressure can be written:

\begin{equation}\label{eos1}
P(\rho,T)=P_c(\rho)+P_i(\rho,T)+P_t(\rho,T),
\end{equation}

where $P_t(\rho,T)=P_e(\rho,T)-P_e(\rho,0)$ is the thermal electronic
contribution to the EOS, $P_e$ being electronic pressure. $P_i$ is the nuclear
(ionic) pressure and $P_c$ the
cold curve. The replacement of $P_e(\rho,0)$ by $P_c(\rho)$ is necessary since
the results of the QSCF models are not valid for $T\rightarrow 0$. Following the
same procedure, internal energy is divided into electronic, ionic and cold
component. Expression (\ref{eos1}) also holds for internal energy. The nuclear
component is evaluated in the ideal-gas approximation.

\begin{figure}
\begin{center}
\includegraphics[width=8.6cm,trim=0 0 0 -60]{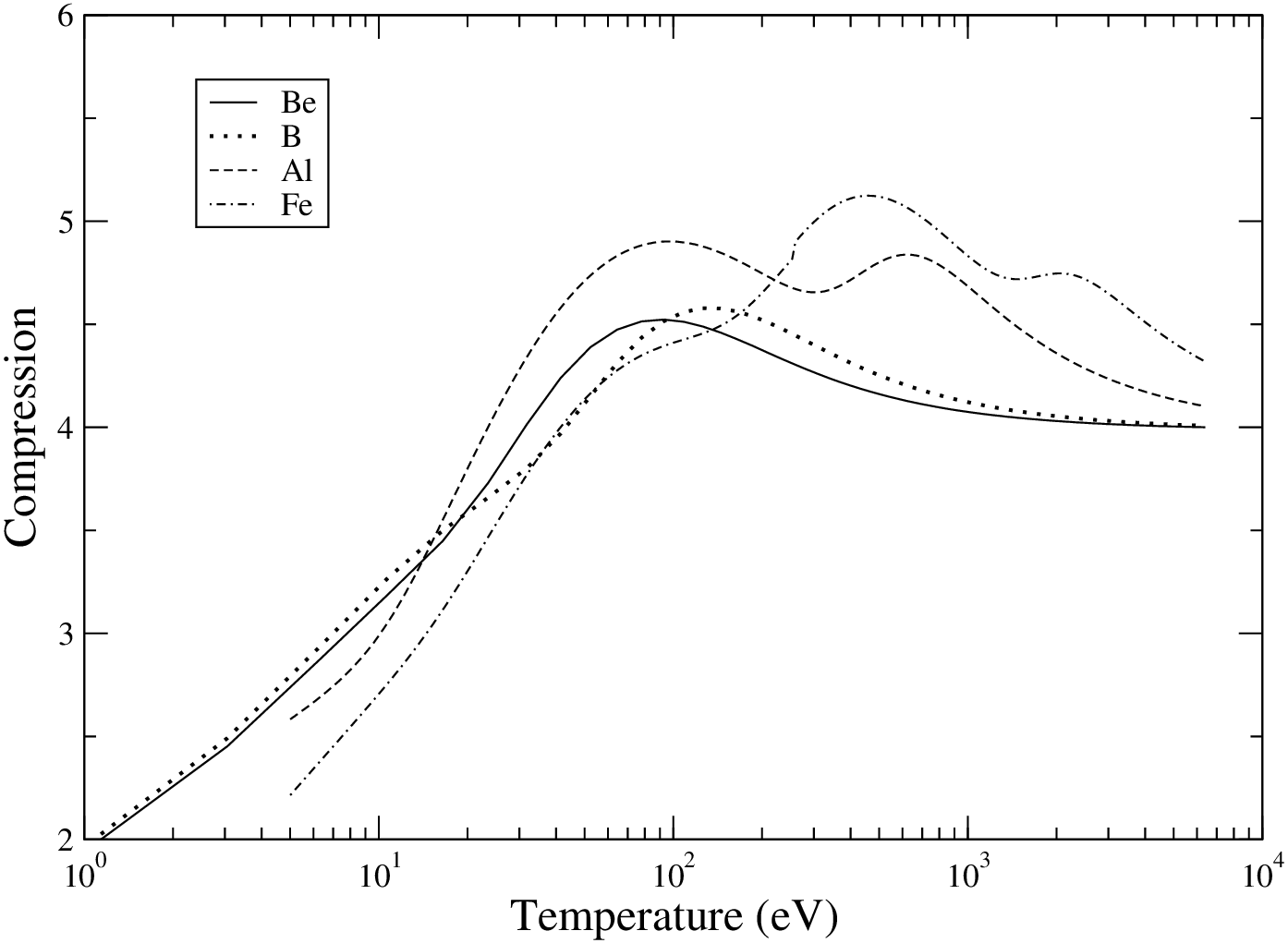}
\end{center}
\caption{Compression versus temperature (thermodynamic path) along the Hugoniot
curve for Be, B, Al and Fe in the ASC model.}
\label{fig8}
\end{figure}

$\bullet$ The cold curve must satisfy the condition that the quantity

\begin{equation}
\rho(\partial P/\partial\rho)
\end{equation}

 evaluated at normal density $\rho_n$ must be
equal to the experimental bulk modulus $B_0$, which can be obtained from the
sound velocity $c_s$ through $B_0=\rho c_s^2$. The cold curves are obtained
either from APW \cite{louc} simulations, or using the Vinet \cite{vin} universal
EOS. In the latter case, pressure can be evaluated analytically by

\begin{equation}
P_c=[3B_0(1-x)/x^2]\exp[\zeta(1-x)]
\end{equation}

and the internal energy by

\begin{equation}
E_c=-\frac{9B_0}{\rho_n\zeta^2}[1-\exp[\zeta(1-x)][1-\zeta(1-x)]]
\end{equation}

with $\zeta=3(B_0'-1)/2$, $x=(\rho/\rho_n)^{1/3}$ and $B_0'=(\partial
B_0/\partial P)|_{P=0,T}$. Values of such parameters obtained from experiments
are available for several elements in the literature (see for instance
\cite{vin,lou}). In many situations, Vinet EOS gives realistic results, and its
accuracy is sufficient for most applications involving high pressures and
temperatures.

\begin{figure}
\begin{center}
\includegraphics[width=8.6cm,trim=0 0 0 -60]{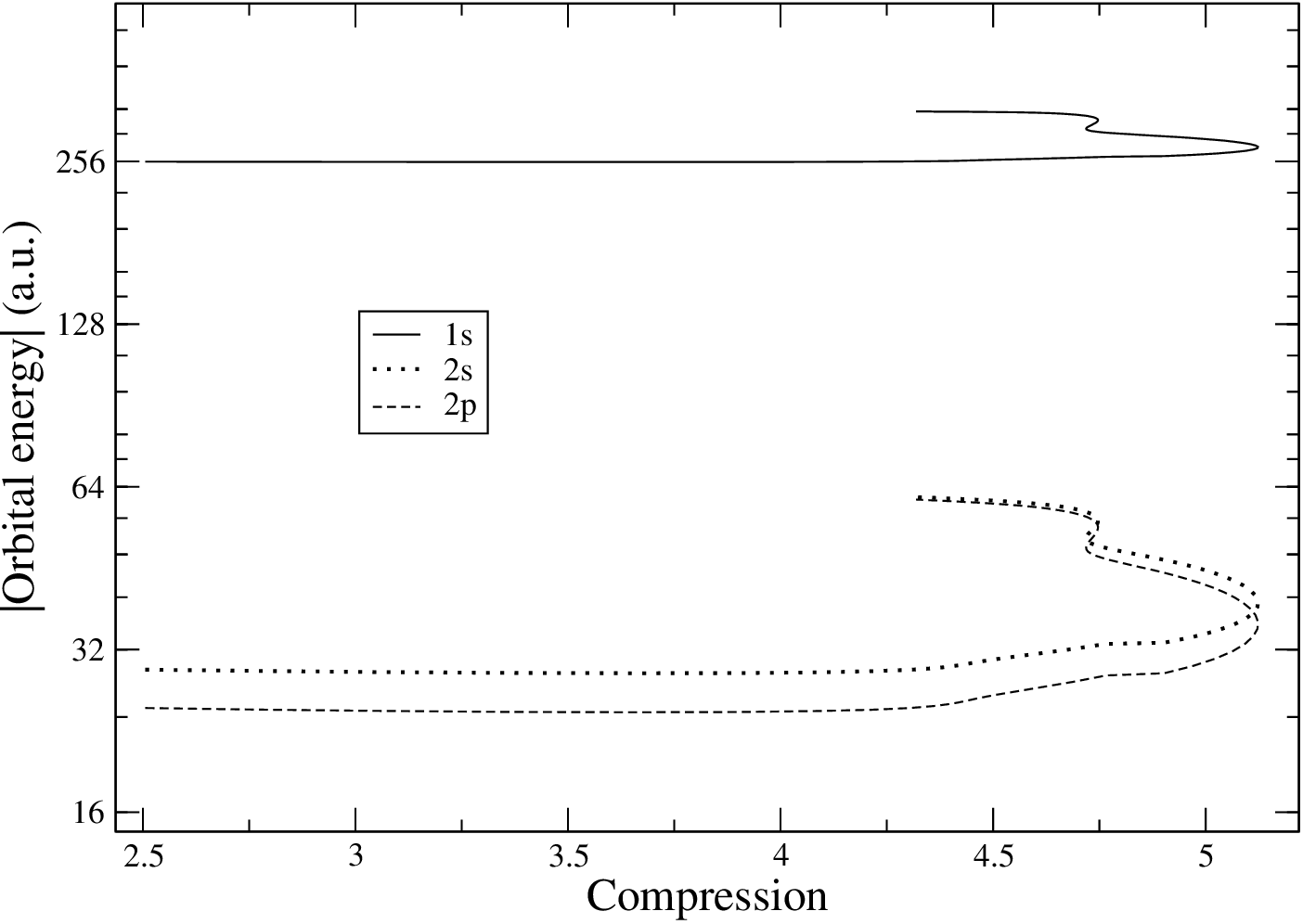}
\end{center}
\caption{Absolute value of eigen-energy of orbitals 1s, 2s and 2p of Fe versus
compression along the Hugoniot in the ASC model. For a given value of $\rho$,
the corresponding value of $T$ can be found on Fig. \ref{fig8}.}
\label{fig9}
\end{figure}


\section{\label{sec3} Shock Hugoniots}


\subsection{\label{subsec31}Definitions}

The initial state of the plasma is characterized by a density $\rho_0$, a
temperature $T_0$, a pressure $P_0$, and an internal energy $E_0$. $D$ is the
shock velocity, $u$ the matter velocity, $P$ and $E$ are respectively the
pressure and internal energy behind the shock front. Since the mass preservation
has to be satisfied, the density of the compressed gas $\rho$ satisfies the
relation $\rho_0D=\rho(D-u)$. The resultant force acting on the compressed gas
is equal to the difference of the pressure on the shock side and on the side of
the undisturbed fluid, that is $P-P_0=\rho_0Du$. The increase in the sum of the
internal and kinetic energies of the compressed gas is equal to the work done by
the external force acting on the shock front, \textit{i.e.}
$\rho_0D(E-E_0+u^2/2)=Pu$. Rearranging these equations, one gets \cite{zel}:

\begin{equation}\label{hug}
\frac{1}{2}(P+P_0)\Big(\frac{1}{\rho_0}-\frac{1}{\rho}\Big)=E-E_0,
\end{equation}

which is known as the Hugoniot relation. It is worth keeping in mind that the
Hugoniot shock adiabat is not the
collection of successive states of the matter during the propagation of the
shock. The only relevant information is that the state of the matter ``after the
shock'' belongs to the Hugoniot. Technically, the Hugoniot curve can be obtained
by solving (\ref{hug}) at each temperature step, either calculating the EOS
``in line'' or by an interpolation in a table (pre-tabulated EOS). The two
relations
$P(\rho,T)$ and $E(\rho,T)$ constitute the EOS, which thermal part is calculated
from the models described in sections \ref{subsec21} and
\ref{subsec22}.

\begin{figure}
\begin{center}
\includegraphics[width=8.6cm,trim=0 0 0 -60]{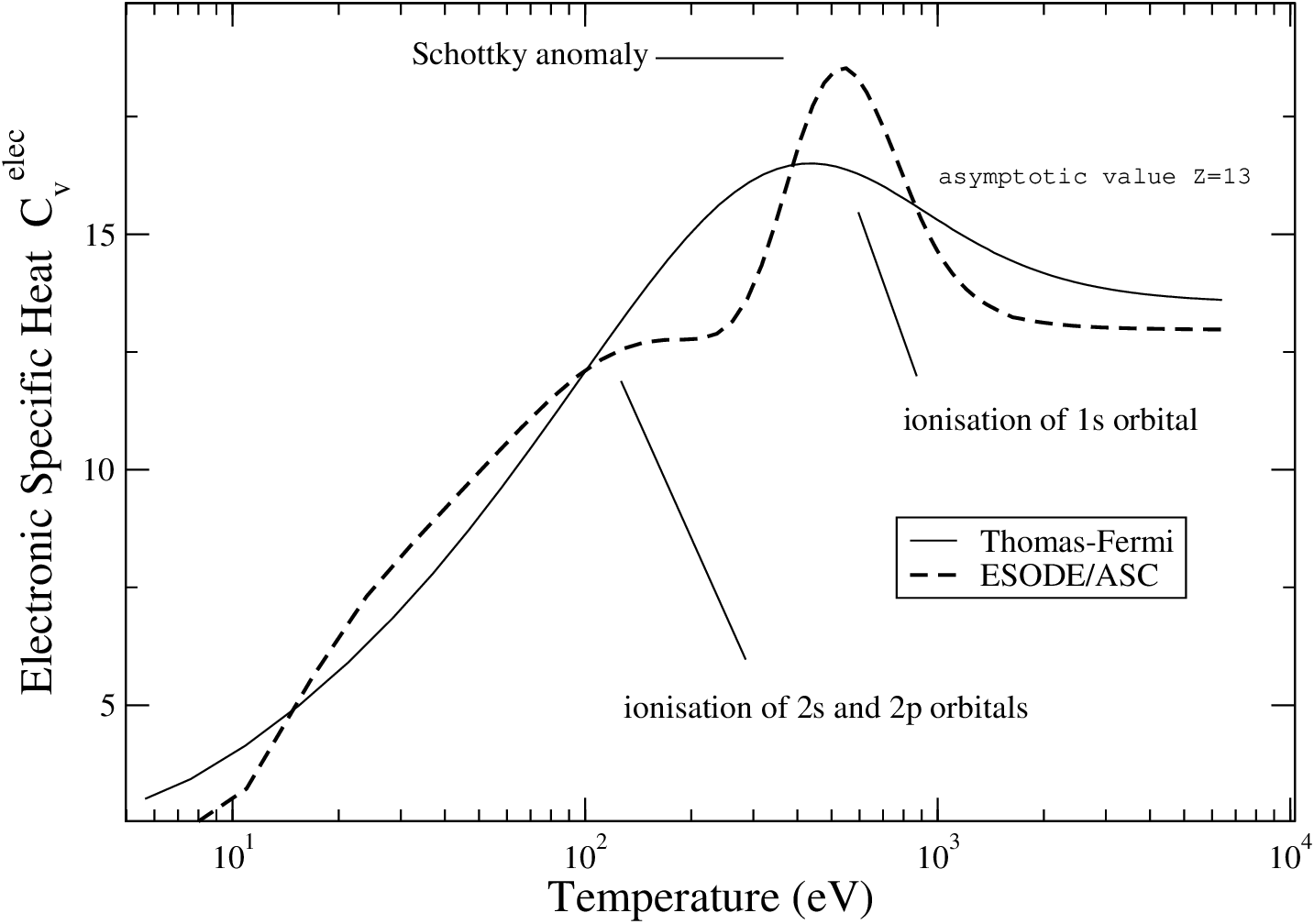}
\end{center}
\caption{Electronic specific heat for Al.}
\label{fig10}
\end{figure}

\begin{figure}
\begin{center}
\includegraphics[width=8.6cm,trim=0 0 0 -60]{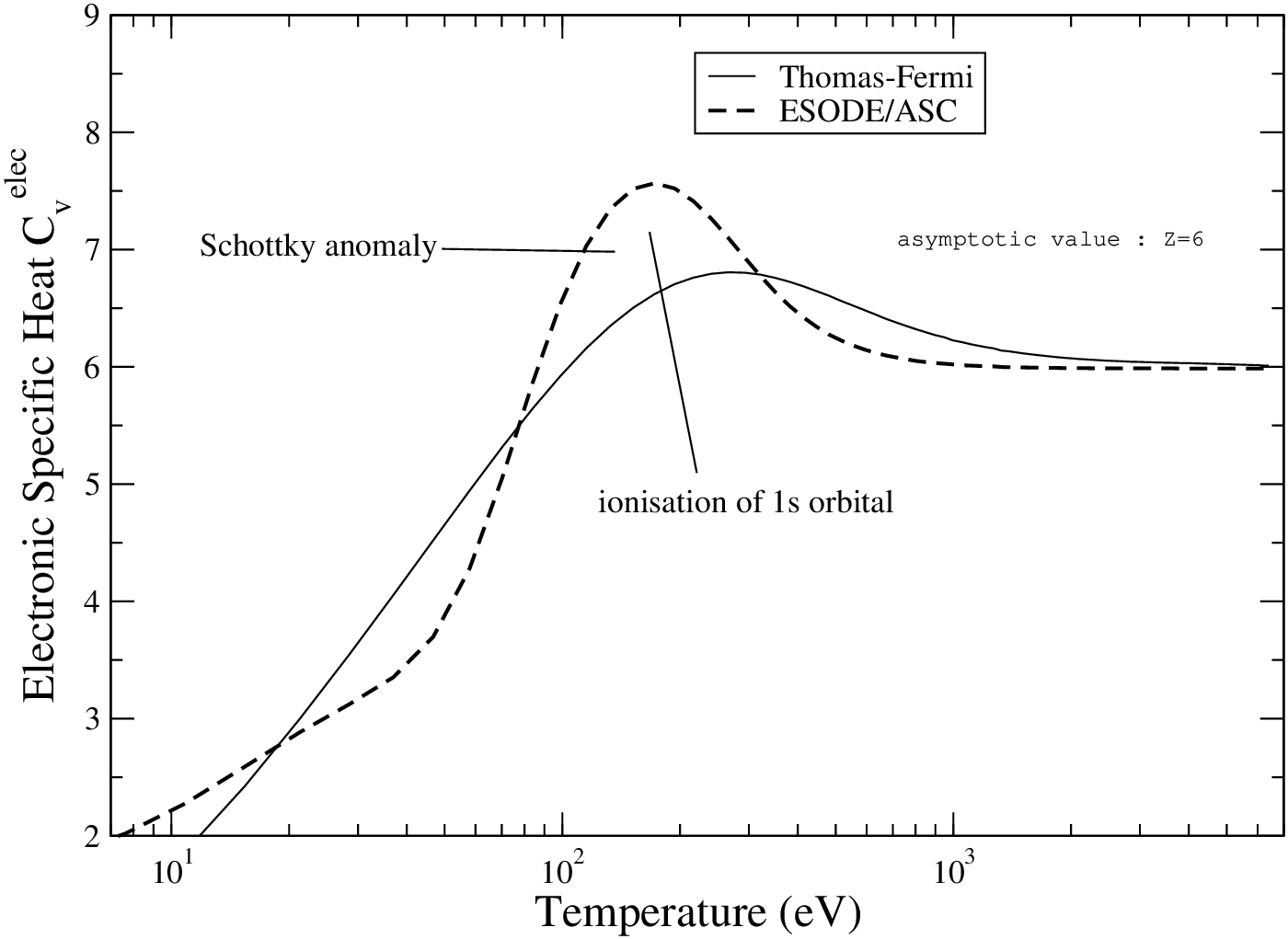}
\end{center}
\caption{Electronic specific heat for C.}
\label{fig11}
\end{figure}

In this work, focus is put on the standard Hugoniot ($P_0$=0, $\rho_0$ equal to
solid density and $T_0$=300 K). No distinction between $\rho_0$ and $\rho_n$
will be made in the following. As mentioned before, total energy is preserved
during the shock, since an increase of internal energy is transformed into heat,
which makes the temperature of the plasma increase. More exactly there is no
heat exchange during the shock (which explains the name ``shock adiabat''), but
an increase of internal energy, distributed in internal degrees of freedom. One
can wonder whether this conservation of energy is in contradiction with the fact
that a physical system tends to minimize energy, as invoked in section
\ref{sec2}.
In classical mechanics, Noether's theorem, which links symmetries and
conservation laws, relates the invariance by translation in time to the
conservation of total energy. Therefore, the idea that a physical system tends
to minimize energy is often included in models where time is absent. This is
also the case of the present study, since in QSCF models the wavefunctions are
assumed to be time-independent. Therefore, only stationary solutions are
considered here, and only systems in thermal equilibrium are supposed to exist.
There is actually no contradiction: total energy is preserved during the shock,
but the pressure calculated in the AJC$_q$ model is thermodynamically consistent
since it is calculated as a derivative of the energy (see (\ref{pv}) in section
\ref{subsec22}). Calculations have been performed for $T\leq$ 6.4 keV. Figures
\ref{fig1}, \ref{fig2}, \ref{fig3}, \ref{fig4}, \ref{fig5} and \ref{fig6}
represent Hugoniot curves for beryllium (Be, $Z$=4), boron (B, $Z$=5), carbon
(C, $Z$=6), aluminum (Al, $Z$=13), iron (Fe, $Z$=26) and copper (Cu, $Z$=29)
respectively, calculated from pure Thomas-Fermi EOS, ASC model, and AJC model
with a full quantum treatment of electrons (namely AJC$_q$). The calculation
within the jellium model with a hybrid description of electrons, \textit{i.e.}
where bound electrons are described in the framework of quantum mechanics and
free electrons in the TF approximation, is also presented (AJC$_h$), since it is
not equivalent to ASC model, even if it relies also on a hybrid description of
electrons. The cold curve has been calculated using APW method for Be, C, Al, Fe
and Cu. For Be, B, Al and Fe, Vinet universal EOS gives, for the Hugoniot curve,
results which we find sufficiently accurate. For B, the Vinet parameters have
been interpolated between Li ($Z$=3) and Be. All the QSCF models give very close
results and differ from the TF approximation; it seems however that both hybrid
models (atom in a spherical cell ASC or in a jellium of charges with quantum
bound electrons and TF free electrons AJC$_h$) are very close to each other and
differ from the full quantum atom-in-a-jellium-of-charges model AJC$_q$. For C,
the Hugoniot curve obtained from ASC model is different from the ones obtained
from other QSCF models, since the K shell is ionized for a lower density in the
ASC approach, which induces a larger pressure. In the case of Al, the difference
between the theories appears for $P\geq$ 3 Mbar and AJC$_q$ gives higher
pressures than the other models for 2$\leq\eta\leq$ 3.5, where
$\eta=\rho/\rho_0$ is the compression rate. For $\eta\geq$ 3.5, all QSCF models
give lower pressures than TF model.

\begin{figure}
\begin{center}
\includegraphics[width=8.6cm,trim=0 0 0 -60]{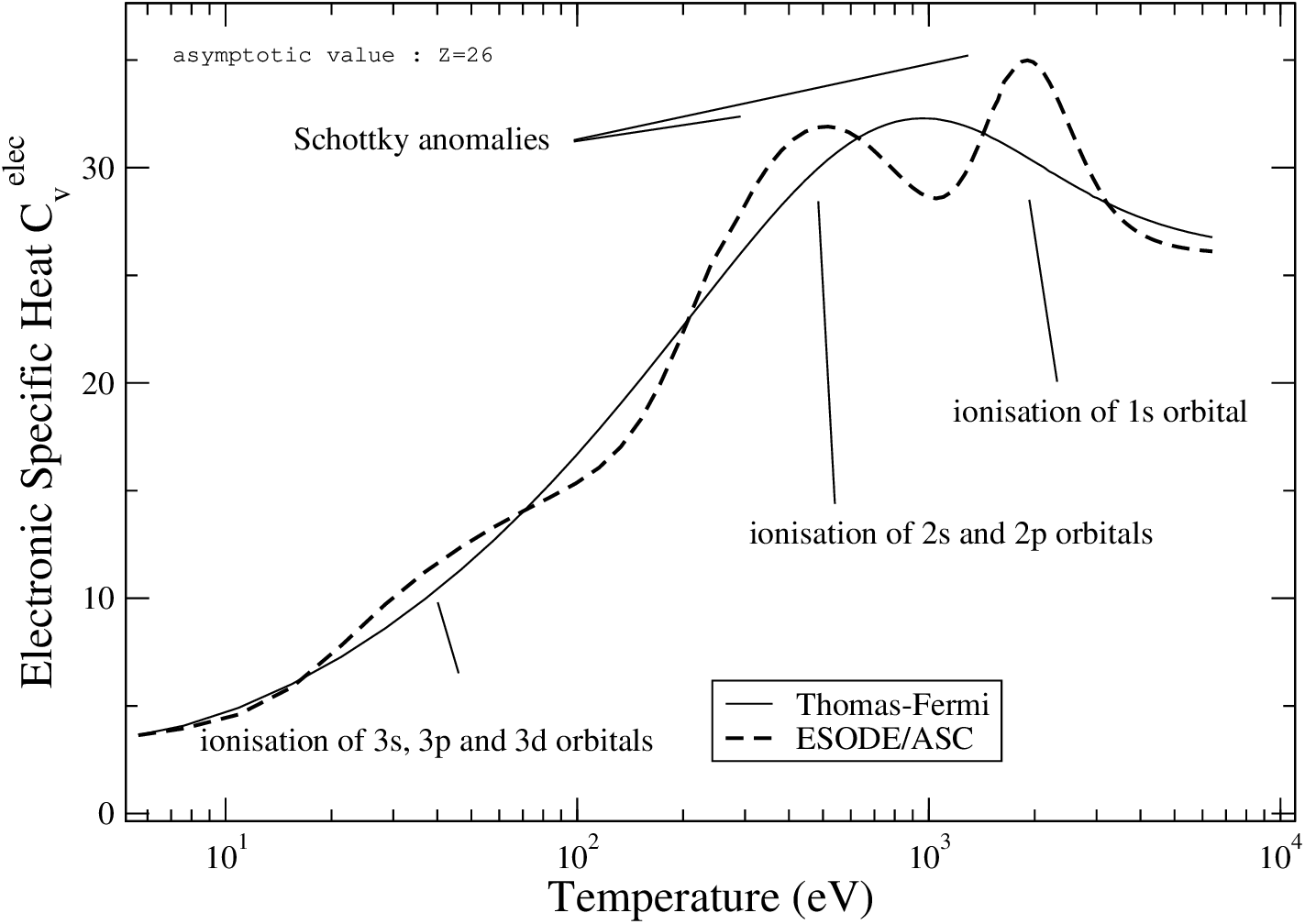}
\end{center}
\caption{Electronic specific heat for Fe.}
\label{fig12}
\end{figure}


\subsection{\label{subsec33}Quantum shell effects}

Our models emphasize the thermodynamic domain where the Hugoniot curve strongly
depends on the electronic structure, \textit{i.e.} beyond four times solid
density where the shoulders (double in the case of Al, Fig. \ref{fig4})
correspond to ionization of successive shells in the Average-Atom picture. The
Hugoniot curve tends to the classical limit equal to four times the solid
density at very high pressure, or typically $T>10$ keV. These shoulders are a
consequence of the competition between the release of energy stocked as internal
energy within the shells and the free-electron pressure. When ionization begins,
the energy of the shock is used mainly to depopulate the relevant shells and the
material is very compressive. However, the pressure of free electrons in
increasing number dominates again and the material becomes more difficult to
compress. Both models show compression maxima in the range $5\rho_0-6\rho_0$. In
this region, the electrons from the ionic cores are being ionized and the shock
density increases beyond the infinite pressure of $4\rho_0$. As ionization is
completed, the plasma approaches an ideal gas of nuclei and electrons and the
density approaches the fourfold density $4\rho_0$. 

\begin{figure}
\begin{center}
\includegraphics[width=8.6cm,trim=0 0 0 -60]{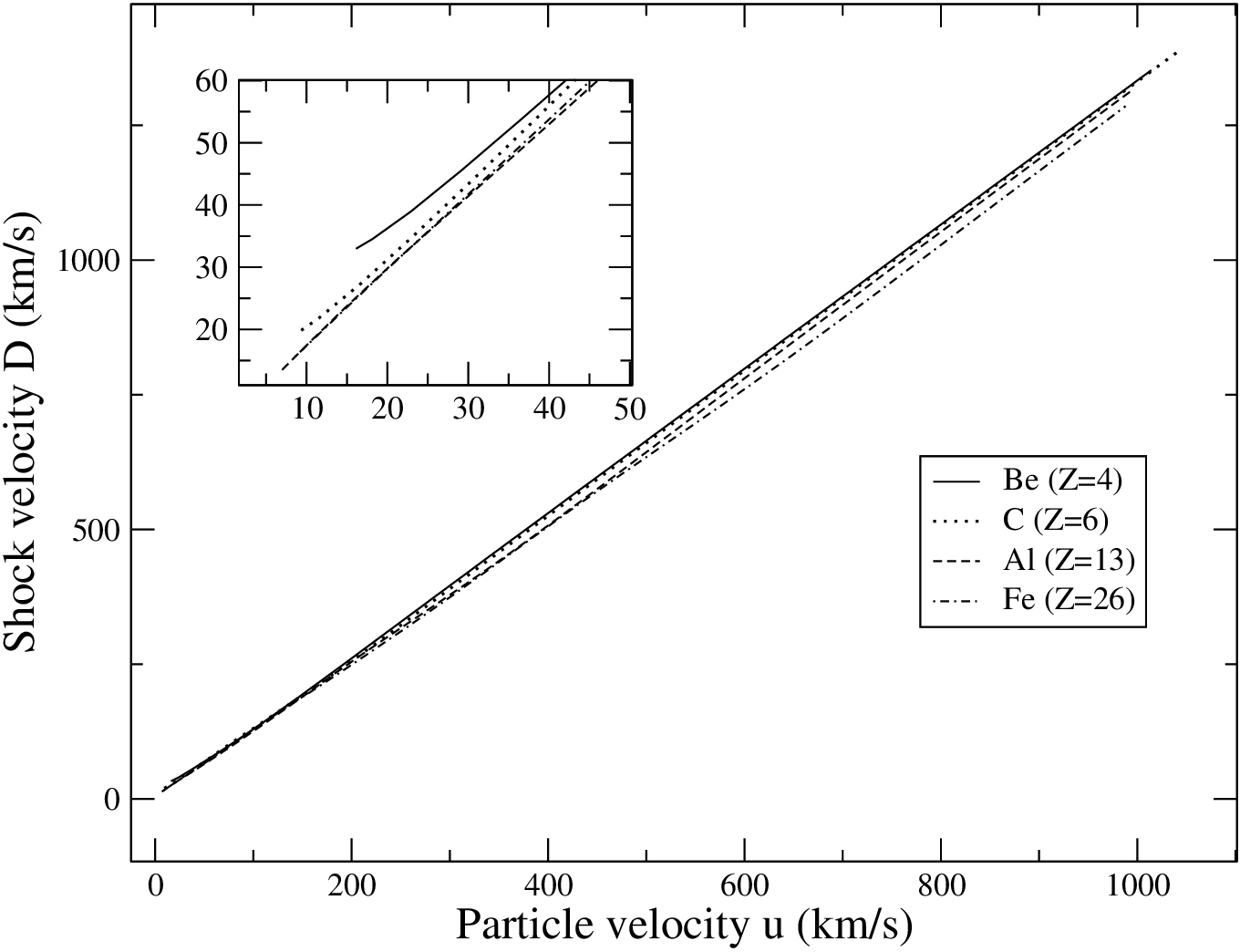}
\end{center}
\caption{Shock velocity $D$ versus particle velocity $u$ for Be, C, Al and Fe.}
\label{fig13}
\end{figure}

The successive electron shells are accurately represented. For Be (Fig.
\ref{fig1}), B (Fig. \ref{fig2}) and C (Fig. \ref{fig3}), all the models
show a
single density maximum, corresponding to the ionization of the K electron shell.
For Al (Fig. \ref{fig4}), there are two density maxima corresponding to the K
and L electron shells. For Fe (Fig. \ref{fig5}) and Cu (Fig. \ref{fig6})
there
are density maxima or inflexions corresponding to the K, L and M electron
shells. The L shell ionization feature gives the largest density increase. In
the case of Cu (Fig. \ref{fig6}), the ASC and AJC$_h$ models exhibit a kind of
discontinuity around 40 g/cm$^3$, due to pressure ionization of $2p$ orbital.
Such a sharp increase of pressure does not exist in the AJC$_q$ model, because
of its thermodynamic consistency. This is due to the fact that it relies on a
variational formulation: pressure is rigorously obtained as a derivative of the
free energy, and shape resonances are carefully taken into account in the
quantum treatment of free electrons. Such features lead to a continuous
disappearing of a bound state into the continuum \cite{koh}. This is not the
case of ASC and AJC$_h$ models. In
the framework of these two models, we tried to improve the dissolution of bound
states into the continuum splitting them into bands and using a simple
degeneracy reduction; this makes the discontinuity a little less sharp but does
not solve the problem. Only a careful search of resonances would help, together
with a variational calculation of pressure. Figure \ref{fig7} illustrates the
fact
that internal energy also reflects the oscillations due to the shell structure,
as well as the thermodynamic path $(T,\rho)$ (see the cases of Be, B, Al and Fe
in Fig. \ref{fig8}). In fact, the non-monotonic character of
thermodynamic variables stems from the eigen-energies of the orbitals
themselves, which exhibit the oscillations as well (see Fig. \ref{fig9}). The
first density for which a lower compression is obtained is named ``turnaround''
point. 

\begin{table}
\begin{center}
\begin{tabular}{|c|c|c|c|c|c|} \hline \hline
 Element & $a_0$ (km/s) & $a_0$ (\cite{alt7}) & $a_1$ & $a_1$ (\cite{alt7}) &
$a_2$ (s/km) \\\hline \hline
 $_4$Be & $9.137$ & $/$ & $1.147$ & $/$ & $5.70 \; 10^{-4}$ \\
 $_5$B & $6.833$ & $/$ & $1.127$ & $/$ & $5.06 \; 10^{-3}$ \\ 
 $_6$C & $10.437$ & $/$ & $0.940$ & $/$ & $5.06 \; 10^{-3}$ \\ 
 $_{11}$Na & $9.718$ & $/$ & $1.063$ & $/$ & $9.58 \; 10^{-4}$ \\ 
 $_{12}$Mg & $4.614$ & $/$ & $1.124$ & $/$ & $7.12 \; 10^{-4}$ \\
 $_{13}$Al & $6.007$ & $6.371$ & $1.152$ & $1.164$ & $4.52 \; 10^{-4}$ \\
 $_{26}$Fe & $6.675$ & $6.982$ & $1.144$ & $1.190$ & $8.42 \; 10^{-4}$ \\
 $_{27}$Co & $5.348$ & $/$ & $1.150$ & $/$ & $7.96 \; 10^{-4}$ \\ 
 $_{29}$Cu & $6.846$ & $5.905$ & $1.137$ & $1.212$ & $6.88 \; 10^{-4}$ \\
 $_{40}$Zr & $6.926$ & $/$ & $1.125$ & $/$ & $3.28 \; 10^{-4}$ \\
 $_{42}$Mo & $6.975$ & $6.711$ & $1.123$ & $1.149$ & $3.50 \; 10^{-4}$ \\
 $_{48}$Cd & $3.676$ & $4.251$ & $1.180$ & $1.182$ & $9.07 \; 10^{-5}$ \\ \hline
\hline 
\end{tabular}
\end{center}
\caption{Shock adiabat parameters of Be, B, C, Na, Mg, Al, Fe, Co, Cu, Zr, Mo
and Cd. Coefficients $a_0$ and $a_1$ are compared with those published for
metals in Ref. \cite{alt7}.
}\label{tab3}
\end{table}

The pressure differences from the quantum mechanical theory along with the
oscillatory feature at the turnaround density, can be explained by examining the
heat capacity predicted by the TF and ASC theories along the Hugoniot path. At
low temperature, the electronic heat capacity depends on the number of electrons
that can be excited around the Fermi energy. The TF theory predicts a smooth
increase since the density of states in this model is a monotonic function of
energy. Therefore, the electronic specific heat
 
\begin{equation}
C_v^{elec}=\frac{1}{3k_B/2}\frac{\partial [E(\rho,T)-E_i(\rho,T)]}{\partial T}
\end{equation}

is an interesting indicator of the ionization of successive shells and
indirectly on the release of energy. Figure \ref{fig10} represents $C_v^{elec}$
along the Hugoniot curve in the TF and in the ASC models for Al. Both theories
show the effect of the coulomb attractive potential of the nucleus binding the
electrons, represented by the peak around 300 eV for TF theory and 100 eV for
ASC model. After the turnaround, there are 11 free electrons and 2 bound
electrons remaining in the $1s$ orbital (K shell), which is far away from the
energy zero (1.5 keV at T=100 eV). This phenomenon is a kind of Schottky anomaly
\cite{sch}. As long as temperature is not sufficient to ionize those two
electrons, the specific heat tends to an asymptote corresponding to an ideal gas
of 11 independent particles. When both $1s$ electrons are ionized (after a
``threshold'' temperature), there is a sudden break in the specific heat, which
tends to an ideal gas of 13 independent particles. It is not as important for
the $2s$ and $2p$ bound states (L shell), since their energy levels are not as
far from the continuum (a few tenth of eV). Figure \ref{fig11} represents the
electronic specific heat along the Hugoniot curve in the TF and in the ASC
models for C. In that case, the only difference between both models is that
there is a strong enhancement due to ionization of $1s$ orbital in the ASC
approach, followed by the asymptotic limit of 6. The same phenomenon occurs with
Fe (Fig. \ref{fig12}); in that case, after two Schottky anomalies, the
electronic part of the specific heat tends to an ideal gas of 26 electrons. 

\begin{figure}
\begin{center}
\includegraphics[width=8.6cm,trim=0 0 0 -60]{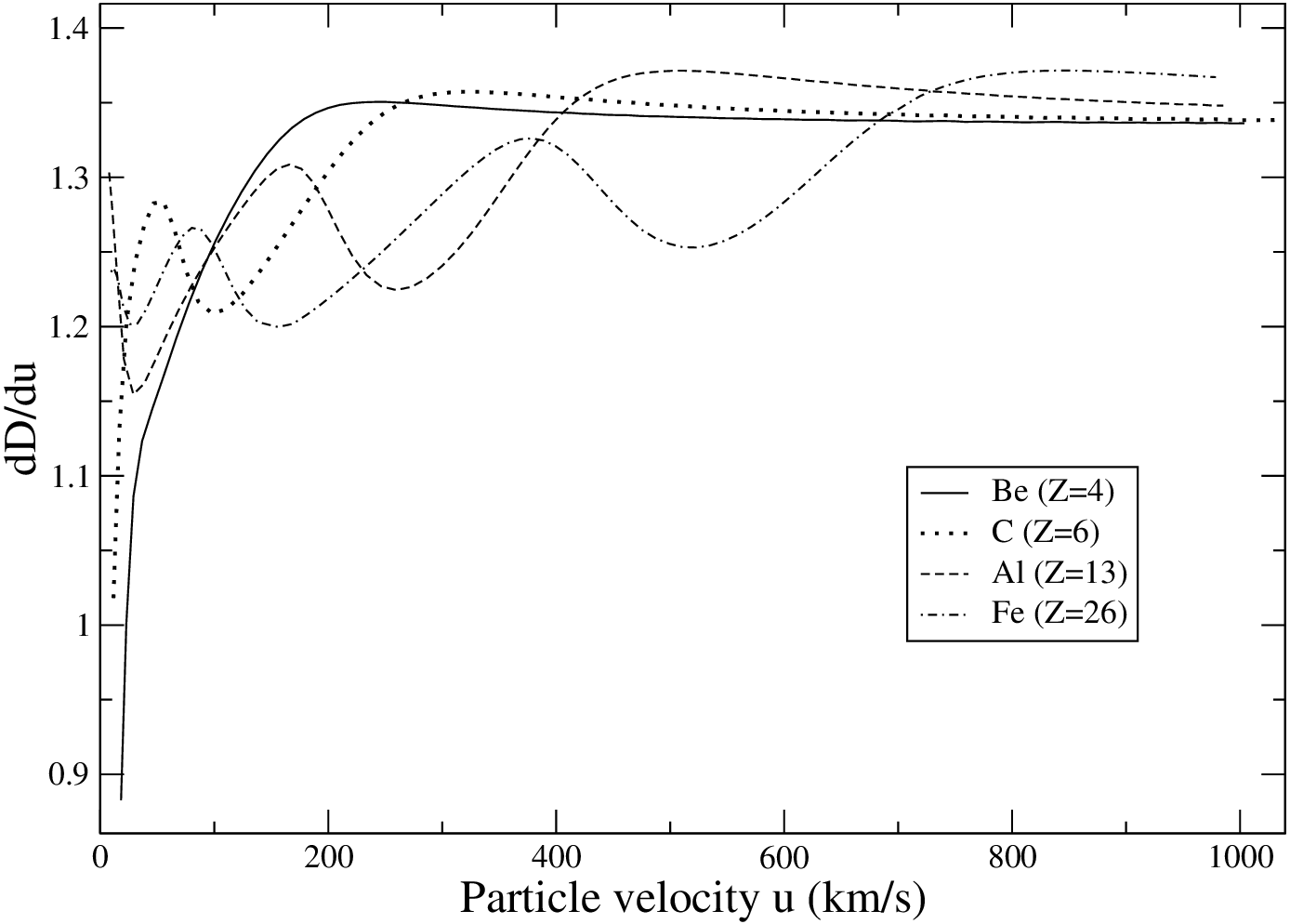}
\end{center}
\caption{Derivative of shock velocity $dD/du$ versus particle velocity $u$ for
Be, C, Al and Fe.}
\label{fig14}
\end{figure} 


\subsection{\label{subsec34}Comparisons with experimental data}\label{tezer}

Experimental data on Be \cite{wal, isb, mar, rag, nel}, B \cite{mar}, C
\cite{mar, gus, nel2}, Al \cite{alt, vol, vla, rag2, sim1, avr, pod, tru2, knu},
Fe \cite{alt, rag2, alt2, kru, alt3, tru3, alt4, tru4, tru5, tru6, alt5} and Cu
\cite{isb, alt, rag2, alt2, tru6, alt5b, kor, tru7, tru8, alt6, glu, mit, tru9}
have been collected for comparison with the EOS models presented above. Several
experimental methods have been used to generate well-defined shock states: gas
guns for pressures up to 5 Mbar, explosive-driven spherical implosions and
laser-driven plane waves for generating shocks up to 10 Mbar. High-power lasers
are capable of driving multimegabar shocks in small samples either by direct
irradiation or indirectly. In the latter case, the target is mounted on a
hohlraum into which the lasers are focused. Pressures up to a few Gbar have been
reached through underground nuclear explosions. If the shocks have been measured
accurately, superimposing all of these data should yield a single smooth shock
Hugoniot curve. The maximum pressures reached in the experiments are: 18 Mbar
for Be, 4000 Mbar for Al, 191 Mbar for Fe and 204 Mbar for Cu. It appears
difficult to decide which model gives the best agreement with experiments,
since there are obviously very few available data for the region of interest
(typically above 100 Mbar). The error bars associated to the ultrahigh-pressure
experimental values for Al are too large to conclude concerning the existence of
shell effects. For Fe and Cu, results from AJC$_q$ model seem to agree better
with the first interpretable experimental points than the hybrid models (ASC or
AJC$_h$). The large errors in the measurements make the experimental results
useless for refining the theoretical data. Therefore, it is not really
legitimate to attempt to relate gas-dynamics measurements with a discussion of
shell effects. Anyway, analysis of the computational results shows that the
deviation from the experimental points of Al can not be explained only by shell
effects. In the region where most of the experimental points are available, the
role of the cold component is important, while shell effects begin to play a
significant role after the matter is already compressed (around the asymptotic
compression $\eta\approx 4$) and begins to heat up. In order to discriminate
models from eachother, the uncertainty on velocities should be very small
(typically $\leq$ 1\%), which is not possible. However, if the temperature
beyond the shock could be measured with an incertainty of about 20 \%, it should
be possible to decide which model gives the best agreement with experimental
data in the $(T,P)$ plane. 

\begin{figure}
\begin{center}
\includegraphics[width=8.6cm,trim=0 0 0 -60]{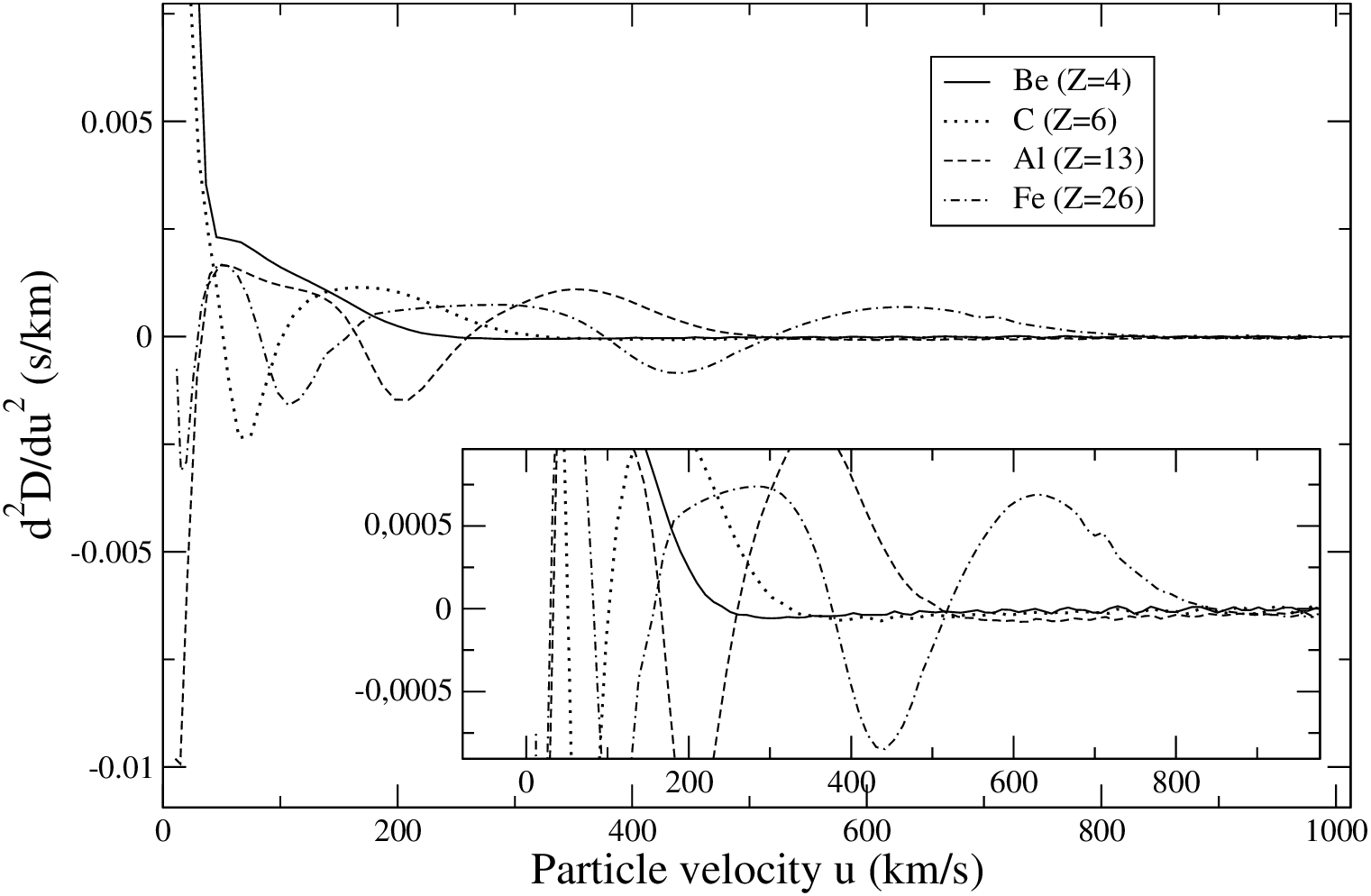}
\end{center}
\caption{Second derivative of shock velocity $d^2D/du^2$ versus particle
velocity $u$ for Be, C, Al and Fe.}
\label{fig15}
\end{figure}


\section{\label{sec4}Shock and particle velocities}

The particle and shock velocities read (see section \ref{subsec31}):

\begin{equation}\label{vel1}
u=\sqrt{\frac{(\rho-\rho_0)(P-P_0)}{\rho_0\rho}} \;\; \mbox{and} \;\;
D=\sqrt{\frac{\rho(P-P_0)}{\rho_0(\rho-\rho_0)}}.
\end{equation}

Expressions in (\ref{vel1}) are generic; they do not involve, \textit{a
priori}, any explicit relation $D(u)$. However, one has \cite{cha}:

\begin{equation}\label{gru}
D=c_s+\frac{\Lambda_0}{2}u+\frac{\Lambda_0}{6c_0}\Big[\frac{G_0}{2}-\frac{1}{
\rho_0}\frac{P_0^{(3)}}{P_0^{(2)}}-\frac{9}{4}\Lambda_0\Big]u^ 2+O(u^3),
\end{equation}

where $G_0$ is the Gr\"uneisen coefficient at $\rho=\rho_0$,

\begin{equation}
\Lambda_0=-P_0^{(2)}/(2\rho_0P_0^{(1)})
\end{equation}

and

\begin{equation}
P^{(n)}=(\partial^{(n)}P/\partial(1/\rho)^{(n)})|_S. 
\end{equation}

\begin{figure}
\begin{center}
\includegraphics[width=8.6cm,trim=0 0 0 -60]{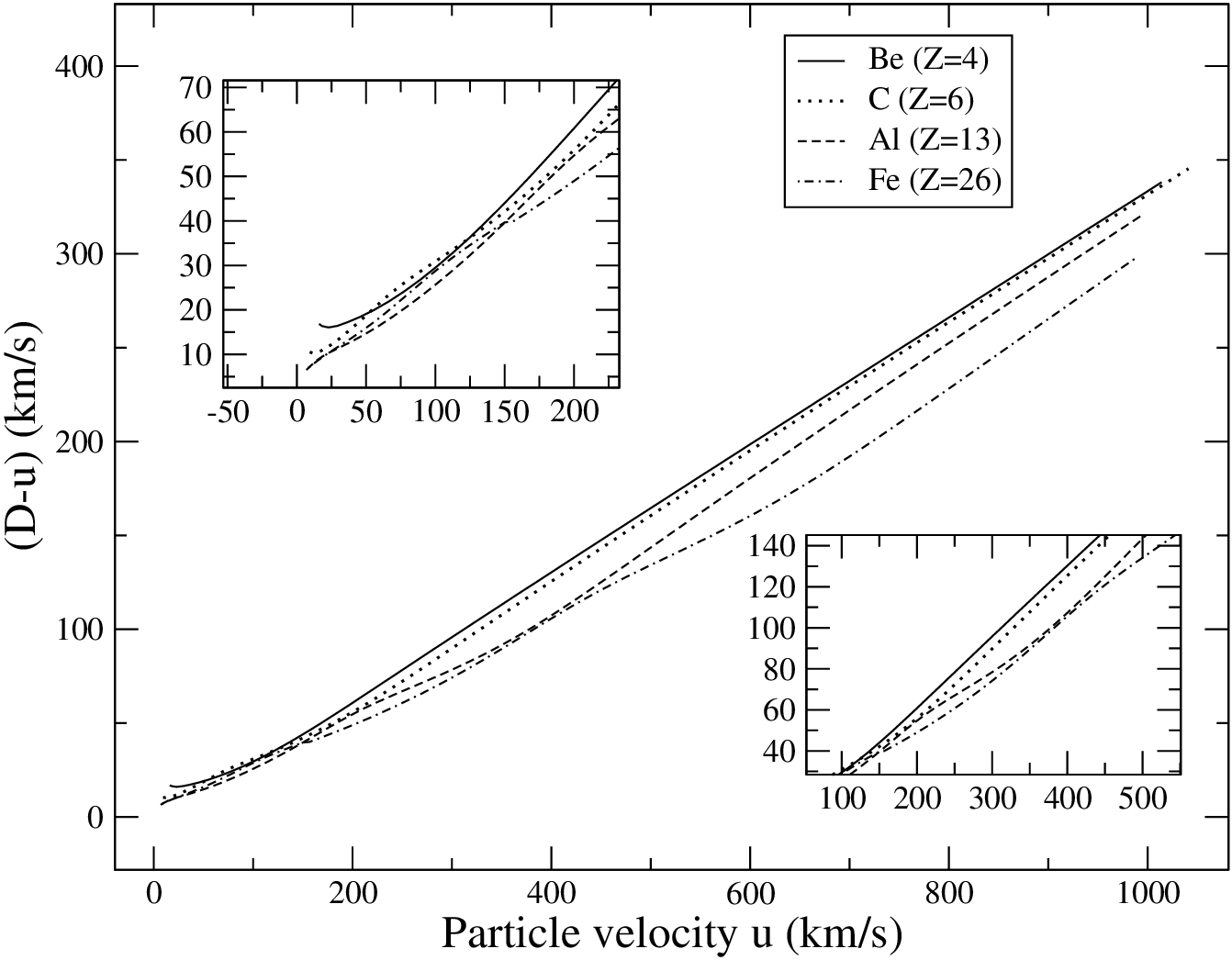}
\end{center}
\caption{$(D-u)$ versus $u$ for Be, C, Al and Fe.}
\label{fig16}
\end{figure}

For a perfect ideal gas,
the Gr\"uneisen parameter $(\partial P/\partial E)|_{\rho}/\rho$ tends to $2/3$.
Expression (\ref{gru}) is suitable for densities close to the pole $(\rho_0,
T_0, P_0,
E_0)$, since it comes from a Taylor development of $D$ and $u$ around that
specific initial point. For metals, it is often invoked that the relation is
quasi-linear. Figure \ref{fig13} illustrates the fact that the $(u,D)$
relationship is almost linear over a wide range of densities, except close to
the origin, where we could not perform the calculation. It is interesting to try
to fit the $(u,D)$ relation with a quadratic expression \cite{alt7}
$D=a_2u^2+a_1u+a_0$. The parameters are presented in table \ref{tab3} for Be, B,
C, Na, Mg, Al, Fe, Co, Cu, Zr, Mo and Cd, and compared to the values published
by Al'tshuler \textit{et al} \cite{alt7}. The slope is different from $4/3$,
corresponding to the perfect ideal gas, as can be checked from (\ref{vel1}).
However, if one looks more carefully at the first and second derivatives (Figs.
\ref{fig14} and \ref{fig15}) of shock velocity versus particle velocity, one
finds that the behaviour of shock velocity is more complicated, and that there
are some oscillations, reflecting the shell structure as well, and even
inflexion points (see Fig. \ref{fig15}). The amplitude of oscillations in the
$(u,D)$ relationship is very small, which is not the case of $(\rho,P)$
relationship. This is even more sensitive at the turnaround of the shock
adiabat. Indeed, in that case, $(\partial P/\partial \rho)|_S$ has strong
variations, which implies, according to the relation

\begin{equation}
\rho^2\frac{\partial P}{\partial\rho}\Big|_S+P\frac{\partial P}{\partial
E}\Big|_{\rho}=\rho^2\frac{\partial P}{\partial \rho}\Big|_E,
\end{equation}

that $P$ has strong variations as well, since $(\partial P/\partial E)|_{\rho}$
and $(\partial P/\partial \rho)|_E$ have smooth variations. A survey of Figs.
\ref{fig13}, \ref{fig14} and \ref{fig15} shows download
curvature, upward curvature and more complicated deviations from non-linear
behaviour. G.I. Kerley \cite{ker} suggests that it is more convenient to plot
$(D-u)$ versus $u$ instead of $D$ versus $u$. His analysis concerns mostly low
velocities, typically below 10 km/s, but Fig. \ref{fig16} shows that it
remains
relevant for high velocities in the region affected by shell effects and
confirms our results.

\section{Conclusion}

The application of shock waves to plasma physics makes possible the generation
under laboratory conditions of extremely high energy densities typical of
astrophysical objects such as stars and giant planets. The physical information
obtained this way extends our basic knowledge on physical properties of plasma
to a broad area of the phase diagram up to pressures nine orders of magnitude
higher than the atmospheric one. 

We presented and compared two equation-of-state models in which the thermal
electronic contribution
is evaluated using two different quantum self-consistent-field models: the
atom
in a spherical
cell (ASC) and the atom in a jellium of charges (AJC). It was shown that both
models are very well suited for the calculation of high-pressure Hugoniot shock
adiabats and that the AJC model, relying on a
full quantum treatment of the electrons, provides a better treatment of pressure
ionization.

We proposed a qualitative and quantitative
study of quantum orbital effects on the principal shock adiabat for different
elements, In the $(\rho,P)$,
$(\rho,E)$ and $(T,\rho)$ representations, such effects
lead to oscillations corresponding to the ionization of successive orbitals.
We found that such oscillations are also visible in the electronic specific
heat, through
Schottky anomalies, and in the energies of the orbitals themselves. They are
responsible for deviations from linear behaviour in the relationship between
shock velocity and particle velocity. 

The next step will be to check
whether the oscillations still exist ``beyond'' the Average Atom model,
\textit{i.e.} if real electronic configurations are taken into
account \cite{pai,pai2}. The remaining difficulty of our models comes
from the impossibility to represent in a simple and suitable way the influence
of the plasma environment on a specific ion. Indeed, such an environment
fluctuates; the number, the localization in space and the structure of
neighbouring ions may change drastically. In principle, it should be necessary
to calculate a large number of geometric configurations of the system (positions
of ions in space), and then their thermodynamic average. Such an approach is
untractable \textit{ab initio}, except for very low-$Z$ atoms in a restricted
range of temperature and density, using molecular-dynamics simulations
\cite{car}. 



\begin{thebibliography}{99}
\bibitem{rem} B.A. Remington, D. Arnett, R.P. Drake and H. Takabe (1999)
\emph{Science}
\textbf{284}, 1488 (1999).

\bibitem{nel0} W.J. Nellis, Contrib. Plasma Phys. \textbf{45}(3-4), 243 (2005).

\bibitem{win} C. Winisdoerffer and G. Chabrier, Phys. Rev. E \textbf{71}, 026402
(2005).

\bibitem{jur1} H. Juranek, V. Schwarz and R. Redmer, J. Phys. A: Math. Gen.
\textbf{36}, 6181 (2003).

\bibitem{jur2} H. Juranek, R. Redmer and W. Stolzmann, Contrib. Plasma Phys.
\textbf{41}, 131 (2001).

\bibitem{cha0} A.Y. Potekhin, G. Massacrier and G. Chabrier, Phys. Rev. E
\textbf{72}, 046402 (2005).

\bibitem{fey} R.P. Feynman, N. Metropolis and E. Teller, Phys. Rev.
\textbf{75},
1561 (1949).

\bibitem{des} M.P. Desjarlais, Contrib. Plasma Phys. \textbf{45}(3-4), 300
(2005).

\bibitem{ebe} W. Ebeling, W. Kraeft and D. Kremp \emph{Theory of Bound States
and Ionization Equilibrium in Plasmas and Solids}, Ackademie-Verlag, Berlin
(1976).

\bibitem{louc} T.L. Loucks \emph{Augmented Plane Wave Method: A Guide to
Performing Electronic Structure Calculations}, W.A. Benjamin, Inc., New York
(1972).

\bibitem{roz2} B.F. Rozsnyai, Phys. Rev. A \textbf{5}, 231 (1972).

\bibitem{pau} H.A. Bethe and E.E. Salpeter \emph{Quantum Mechanics of One- and
Two-electron Atoms}, Springer Verlag, Berlin (1957).

\bibitem{iye} H. Iyetomi and S. Ichimaru, Phys. Rev. A \textbf{34}, 433 (1986).

\bibitem{pai} J.C. Pain, G. Dejonghe and T. Blenski, J. Quant. Spectrosc.
Radiat.
Transfer \textbf{99}, 451 (2006).

\bibitem{pai2} J.C. Pain, G. Dejonghe and T. Blenski, J. Phys. A: Math.
Gen. \textbf{39}, 4659 (2006).

\bibitem{fri1} J. Friedel, Philos. Mag. \textbf{43}, 153 (1952).

\bibitem{fri2} J Friedel, Adv. Phys. \textbf{3}, 446 (1954).

\bibitem{dag1} L. Dagens, J. Phys. C \textbf{5}, 2333 (1972).

\bibitem{dag2} L. Dagens, J. Phys. (Paris) \textbf{34}, 879 (1973).

\bibitem{per1} F. Perrot, Phys. Rev. A \textbf{42}, 4871 (1990).

\bibitem{per2} F. Perrot, Phys. Rev. E \textbf{47}, 570 (1993).

\bibitem{per3} F. Perrot, \emph{Recherche d'un mod\`ele de structure
\'electronique
des plasmas applicable aux calculs d'opacit\'e et d'\'equation d'\'etat} (1998),
unpublished.

\bibitem{lib} D.A. Liberman, Phys. Rev. B \textbf{20}, 4981 (1979).

\bibitem{roz1} B.F. Rozsnyai, J.R. Albritton, D.A. Young, V.N. Sonnad and D.A.
Liberman, Phys. Lett. A \textbf{291}, 226 (2001).

\bibitem{mor2} R.M. More \textit{Atomic Physics in Inertial Confinement
Fusion}, UCRL
Report 84991 (1981).

\bibitem{sal} Salzmann D \emph{Atomic Physics in Hot Plasmas}, International
series of monographs on physics, New York, Oxford University Press, p 28 (1998).

\bibitem{vin} P. Vinet, J.H. Rose, J. Ferrante and J.R. Smith, J.Phys.:
Condens.
Matter \textbf{1}, 1941 (1989).

\bibitem{lou} A. Dewaele, P. Loubeyre and M. Mezouar, Phys. Rev. B \textbf{70},
094112 (2004).

\bibitem{zel} Ya.B. Zel'dovich and Yu.P. Raizer, \emph{Physics of Shock
Waves
and
High-Temperature Hydrodynamic Phenomena}, Academic Press, New York, Vol
1, Ch 3 (1966).

\bibitem{koh} W. Kohn and C. Majundar, Phys. Rev. \textbf{140}, A1133 (1965).

\bibitem{sch} W. Schottky, Physik. Z. \textbf{23}, 448 (1922).

\bibitem{rus} A.V. Bushman, I.V. Lomonosov, K.V. Khishchenko, \emph{Shock Wave
Database}
and references therein (http://teos.ficp.ac.ru/rusbank/).

\bibitem{wal} J.M. Walsh, M.H. Rice, R.G. McQueen, F.L. Yarger, Phys. Rev.
\textbf{108}, 196 (1957). 

\bibitem{isb} W.H. Isbell, F.H. Shipman, A.H. Jones, \textit{Hugoniot equation
of state measurements for eleven materials to five megabars}, General Motors
Corp. Materials Science Laboratory Report MSL-68-13 (1968). 

\bibitem{mar} S.P. March, LASL Hugoniot Data, University of California Press,
Berkeley, 1980. 

\bibitem{rag} C.E. Ragan, Phys. Rev. Ser. A \textbf{25}, 3360 (1982). 

\bibitem{nel} W.J. Nellis, J.A. Moriarty, A.C. Mitchell, N.C. Holmes, J. Appl.
Phys. \textbf{82}, 2225 (1997). 

\bibitem{gus} W.H. Gust, Phys. Rev. B \textbf{22}, 4744 (1980). 

\bibitem{nel2} W.J. Nellis, A.C. Mitchell, A.K. McMahan, J. Appl. Phys.
\textbf{90}, 696 (2001). 

\bibitem{alt} L.V. Al'tshuler, N.N. Kalitkin, L.V. Kuz'mina, B.S. Chekin, Zh.
Eksp. Teor. Fiz. \textbf{72}, 317 (1977), Sov. Phys. JETP \textbf{45}, 167
(1977). 

\bibitem{vol} L.P. Volkov, N.P. Voloshin, A.S. Vladimirov, V.N. Nogin, V.A.
Simonenko, Pis'ma Zh. Eksp. Teor. Fiz. \textbf{31}, 623 (1980), JETP Lett.
\textbf{31}, 588 (1980). 

\bibitem{vla} A.S. Vladimirov, N.P. Voloshin, V.N. Nogin, A.V. Petrovtsev, V.A.
Simonenko, Pis'ma Zh. Eksp. Teor. Fiz. \textbf{39}, 69 (1984), JETP Lett.
\textbf{39}, 85 (1984). 

\bibitem{rag2} C.E. Ragan, Phys. Rev. Ser. A \textbf{29}, 1391 (1984). 

\bibitem{sim1} V.A. Simonenko, N.P. Voloshin, A.S. Vladimirov, A.P. Nagibin,
V.N. Nogin, V.A. Popov, V.A. Vasilenko, Yu.A. Shoidin, Zh. Eksp. Teor. Fiz.
\textbf{88}, 1452 (1985), Sov. Phys. JETP \textbf{61}, 869 (1985). 

\bibitem{avr} E.N. Avrorin, B.K. Vodolaga, N.P. Voloshin, V.F. Kuropatenko, G.V.
Kovalenko, V.A. Simonenko, B.T. Chernodolyuk, Pis'ma Zh. Eksp. Teor. Fiz.
\textbf{43}, 241 (1986), JETP Lett. \textbf{43}, 309 (1986). 

\bibitem{pod} M.A. Podurets, V.M. Ktitorov, R.F. Trunin, L.V. Popov, A.Ya.
Matveev, B.V. Pechenkin, A.G. Sevast'yanov, Teplofiz. Vys. Temp. \textbf{32},
952 (1994). 

\bibitem{tru2} R.F. Trunin, N.V. Panov, A.B. Medvedev, Pis'ma Zh. Eksp. Teor.
Fiz. \textbf{62}, 572 (1995). 

\bibitem{knu} M.D. Knudson, R.W. Lemke, D.B. Hayes, C.A. Hall, C. Deeney, J.R.
Asay, J. Appl. Phys. \textbf{94}, 4420 (2003). 

\bibitem{alt2} L.V. Al'tshuler, A.A. Bakanova, R.F. Trunin, Zh. Eksp. Teor. Fiz.
\textbf{42}, 91 (1962), Sov. Phys. JETP \textbf{15}, 65 (1962). 

\bibitem{kru} K.K. Krupnikov, A.A. Bakanova, M.I. Brazhnik, R.F. Trunin, Dokl.
Akad. Nauk SSSR \textbf{148}, 1302 (1963), Sov. Phys. Dokl. \textbf{8}, 205
(1963). 

\bibitem{alt3} L.V. Al'tshuler, B.N. Moiseev, L.V. Popov, G.V. Simakov, R.F.
Trunin, Zh. Eksp. Teor. Fiz. \textbf{54}, 785 (1968), Sov. Phys. JETP
\textbf{27}, 420 (1968). 

\bibitem{tru3} R.F. Trunin, M.A. Podurets, G.V. Simakov, L.V. Popov, B.N.
Moiseev, Zh. Eksp. Teor. Fiz. \textbf{62}, 1043 (1972), Sov. Phys. JETP
\textbf{35}, 550 (1972). 

\bibitem{alt4} L.V. Al'tshuler, A.A. Bakanova, I.P. Dudoladov, E.A. Dynin, R.F.
Trunin, B.S. Chekin, Zh. Prikl. Mekh. Tekhn. Fiz. \textbf{2}, 3 (1981), J. Appl.
Mech. Techn. Phys. \textbf{22}, 145 (1981). 

\bibitem{tru4} R.F. Trunin, M.A. Podurets, L.V. Popov, V.N. Zubarev, A.A.
Bakanova, V.M. Ktitorov, A.G. Sevast'yanov, G.V. Simakov, I.P. Dudoladov, Zh.
Eksp. Teor. Fiz. \textbf{102}, 1433 (1992), Sov. Phys. JETP \textbf{75}, 777
(1992). 
 
\bibitem{tru5} R.F. Trunin, M.A. Podurets, L.V. Popov, B.N. Moiseev, G.V.
Simakov, A.G. Sevast'yanov, Zh. Eksp. Teor. Fiz. \textbf{103}, 2189 (1993), JETP
\textbf{76}, 1095 (1993). 
 
\bibitem{tru6} R.F. Trunin, Usp. Fiz. Nauk \textbf{164}, 1215 (1994). 

\bibitem{alt5} L.V. Al'tshuler, R.F. Trunin, K.K. Krupnikov, N.V. Panov, Usp.
Fiz. Nauk \textbf{164}, 575 (1996), Sov. Phys. Usp. \textbf{39}, 539 (1996).
  
\bibitem{alt5b} L.V. Al'tshuler, S.B. Kormer, A.A. Bakanova, R.F. Trunin, Zh.
Eksp. Teor. Fiz. \textbf{38}, 790 (1960), Sov. Phys. JETP \textbf{11}, 573
(1960). 

\bibitem{kor} S.B. Kormer, A.I. Funtikov, V.D. Urlin, A.N. Kolesnikova, Zh.
Eksp. Teor. Fiz. \textbf{42}, 686 (1962), Sov. Phys. JETP \textbf{15}, 477
(1962). 
      
\bibitem{tru7} R.F. Trunin, M.A. Podurets, B.N. Moiseev, G.V. Simakov, L.V.
Popov, Zh. Eksp. Teor. Fiz. \textbf{56}, 1172 (1969), Sov. Phys. JETP
\textbf{29}, 630 (1969). 
    
\bibitem{tru8} R.F. Trunin, M.A. Podurets, G.V. Simakov, L.V. Popov, B.N.
Moiseev, Zh. Eksp. Teor. Fiz. \textbf{62}, 1043 (1972), Sov. Phys. JETP
\textbf{35}, 550 (1972). 
    
\bibitem{alt6} L.V. Al'tshuler, B.S. Chekin, \textit{Metrology of high pulsed
pressures}, in: Proceed. of 1st All-Union Pulsed Pressures Simposium (VNIIFTRI,
Moscow, 1974) \textbf{1}, 5. 
            
\bibitem{glu} B.L. Glushak, A.P. Zharkov, M.V. Zhernokletov, V.Ya. Ternovoi,
A.S. Filimonov, V.E. Fortov, Zh. Eksp. Teor. Fiz. \textbf{96}, 1301 (1989), Sov.
Phys. JETP \textbf{69}, 739 (1989). 
    
\bibitem{mit} A.C. Mitchell, W.J. Nellis, J.A. Moriarty, R.A. Heinle, N.C.
Holmes, R.E. Tipton, G.W. Repp, J. Appl. Phys. \textbf{69}, 2981 (1991). 

\bibitem{tru9} R.F. Trunin, L.A. Il'kaeva, M.A. Podurets, L.V. Popov, B.V.
Pechenkin, L.V. Prokhorov, A.G. Sevast'yanov, V.V. Khrustalev, Teplofiz. Vys.
Temp. \textbf{32}, 692 (1994). 

\bibitem{cha} F. Chaiss\'e \emph{Propri\'et\'es structurales de la courbe
d'Hugoniot}, CEA Report R-6013 (2002). 

\bibitem{alt7} L.V. Al'tshuler, R.F. Trunin, V.D. Urlin, V.E. Fortov and V.E.
Funtikov, Sov. Phys. Usp. \textbf{42},
261 (1999).

\bibitem{ker} G.I. Kerley, \emph {The linear $U_s-u_p$ Relation in
Shock-Wave
Physics}, Research Report KTS06-1, Kerley Technical Services, P.O. Box 709,
Appomattox, VA 24522-0709) (2006).

\bibitem{car} R. Car and M. Parrinello, Phys. Rev. Lett., \textbf{55},
2471 (1985).

\end{thebibliography}
\end{document}